\documentclass[twocolumn,showpacs,superscriptaddress,prb]{revtex4-1}
\usepackage{graphicx}
\usepackage{caption}
\usepackage{subcaption}
\RequirePackage[font=footnotesize,labelfont={bf,it}]{caption}
\usepackage[T1]{fontenc}
\usepackage{textcomp}
\usepackage{amsmath,amssymb}
\usepackage{tikz}
\usepackage{soul}
\usepackage{color}
\usetikzlibrary{positioning,arrows}
\usetikzlibrary{decorations.pathmorphing}
\usetikzlibrary{decorations.markings}
\definecolor{myblue}{RGB}{56,94,141}

\begin{document}
\tikzset{gluon/.style={decorate,draw=red,decoration={coil,aspect=0}},
photon/.style={decorate,decoration={snake=bumps},draw=blue}
bracing/.style={decorate,decoration=brace,draw=myblue}
crossing/.style= {decorate,decoration=zigzag,pre=crosses,pre length=0.5cm,draw=myblue}
}
\newcommand{\newc}{\newcommand}

\newc{\kt}{\rangle}
\newc{\br}{\langle}

\newc{\pr}{\prime}
\newc{\longra}{\longrightarrow}
\newc{\ot}{\otimes}
\newc{\rarrow}{\rightarrow}
\newc{\h}{\hat}
\newc{\bom}{\boldmath}
\newc{\btd}{\bigtriangledown}
\newc{\al}{\alpha}
\newc{\be}{\beta}
\newc{\ld}{\lambda}
\newc{\sg}{\sigma}
\newc{\p}{\psi}
\newc{\eps}{\epsilon}
\newc{\om}{\omega}
\newc{\mb}{\mbox}
\newc{\tm}{\times}
\newc{\hu}{\hat{u}}
\newc{\hv}{\hat{v}}
\newc{\hk}{\hat{K}}
\newc{\ra}{\rightarrow}
\newc{\non}{\nonumber}

\newc{\dg}{\dagger}
\newc{\prh}{\mbox{PR}_H}
\newc{\prq}{\mbox{PR}_q}
\newc{\tr}{\mbox{tr}}
\newc{\pd}{\partial}
\newc{\qv}{\vec{q}}
\newc{\pv}{\vec{p}}
\newc{\dqv}{\delta\vec{q}}
\newc{\dpv}{\delta\vec{p}}
\newc{\mbq}{\mathbf{q}}
\newc{\mbqp}{\mathbf{q'}}
\newc{\mbpp}{\mathbf{p'}}
\newc{\mbp}{\mathbf{p}}
\newc{\mbn}{\mathbf{\nabla}}
\newc{\dmbq}{\delta \mbq}
\newc{\dmbp}{\delta \mbp}
\newc{\T}{\mathsf{T}}
\newc{\J}{\mathsf{J}}
\newc{\sfL}{\mathsf{L}}
\newc{\C}{\mathsf{C}}
\newc{\B}{\mathsf{M}}
\newc{\V}{\mathsf{V}}
\title{Generation of coherence in an exactly solvable nonlinear nanomechanical system}
\author{A. K. Singh}
\email{abhishekkrsingh.rs.phy17@iitbhu.ac.in}
\affiliation{ Department of Physics, Indian Institute of Technology (Banaras Hindu University) Varanasi - 221005, India}
\author{L.~Chotorlishvili}
\affiliation{Institut f\"ur Physik, Martin-Luther-Universit\"at Halle-Wittenberg, D-06099 Halle, Germany}
\author{S. Srivastava}
\affiliation{ Department of Physics, Indian Institute of Technology (Banaras Hindu University) Varanasi - 221005, India}
\author{I.~Tralle}
\affiliation{Faculty of Mathematics and Natural Sciences, University of Rzeszow,   Pigonia str. 1, 35-310 Rzeszow, Poland}
\author{Z.~Toklikishvili}
\affiliation{Faculty of Mathematics and Natural Sciences, Tbilisi State University, Chavchavadze av.3, 0128 Tbilisi, Georgia}
\author{ J.~Berakdar}
\affiliation{Institut f\"ur Physik, Martin-Luther-Universit\"at Halle-Wittenberg, D-06099 Halle, Germany}
\author{S. K. Mishra}
\email{sunilkm.app@iitbhu.ac.in}
\affiliation{ Department of Physics, Indian Institute of Technology (Banaras Hindu University) Varanasi - 221005, India}

\begin{abstract}
This study is focused on the quantum dynamics of a nitrogen-vacancy (NV) center coupled to a nonlinear, periodically driven mechanical oscillator.  For  a continuous periodic  driving that depends  on the position of the oscillator, the mechanical motion is described by   Mathieu elliptic  functions.  This solution is employed to study   the dynamics of the quantum spin system including environmental effects and to evaluate the purity and the von Neumann entropy of the NV-spin. The  unitary generation of coherence is addressed. We observe that the production of coherence through a unitary transformation depends on whether the system is prepared initially in mixed state. Production of coherence is efficient when the system initially is prepared in the region of the separatrix (i.e., the region where classical systems exhibit dynamical chaos). From the theory of dynamical chaos, we know that phase trajectories of the system passing through the homoclinic tangle have limited memory, and therefore the information about the initial conditions is lost. We proved that quantum chaos and diminishing of information about the mixed initial state favors the generation of quantum coherence through the unitary evolution. We introduced quantum distance from the homoclinic tangle and proved that for the initial states permitting efficient generation of coherence,  this distance is minimal.
\end{abstract}
\date{\today}
\maketitle
\section{Introduction}
Experimental advances in fabrication and characterization of a nano-electromechanical systems (NEMS), quantum opto-electromechanics, cavity quantum electrodynamics gave further impetus to the  fields of quantum computation and quantum hybrid systems.
\cite{Naik2009,Connell2010,Alegre2011,Stannigel2010,Safavi-Naeini2011,Camerer2011,Eichenfield2009,Safavi-Naeini2012,Brahms2012,Nunnenkamp2012,Khalili2012,Meaney2011,Atalaya2011,Rabl2010,Prants2011,Ludwig2010,Schmidt2010,Karabalin2009,Chotorlishvili_2011,Shevchenko2012,Liu2010,Shevchenko2010,Zueco2009,Cohen2013,Rabl2009,Zhou2010,Chotorlishvili2013}
NEMS as hybrid systems are important for  quantum information transfer, and to facilitate entanglement \cite{Liu2016}, and also serve for studying fundamental questions at the quantum-classical boundaries.
Key features of NEMS are the high Q-factors, low masses and the high frequency of the mechanical oscillations (of the order of Gigahertz (GHz)) \cite{Gaidarzhy2007}.  Recently,   entangling  two micro-mechanical oscillators has been achieved \cite{Ockeloen-Korppi2018}.
Efficient experimental implementation of quantum control has also been achieved using a quantum opto-electro-mechanical protocol\cite{Rogers2014}. With ground-state cooling \cite{Rocheleau2010, Verhagen2012}  exploring the  quantum nature of the mechanical motion becomes feasible. Furthermore, the coupling of a nanomechanical resonator to a nearby (quantum) spin  was studied \cite{Arcizet2011}.
In addition, on the basis of these  hybrid systems various realizations of qubits were proposed and realized. One such system is a NV-center, which is a nitrogen vacancy defect in a diamond lattice. The researchers in this area are mainly interested in the dynamics of the NV center that can be described effectively by  a spin-1 system with a large decoherence time. The Hamiltonian of the NV center couples the ground state $|0\rangle$ to a bright superposition of excited states $|b\rangle=\frac{1}{\sqrt{2}}\left(|-1\rangle+|1\rangle\right)$, while the ``dark'' superposition $|d\rangle=\frac{1}{\sqrt{2}}\left(|-1\rangle-|1\rangle\right)$ remains decoupled
\cite{Rabl2009}. This allows to map the NV centers in external microwave driving onto a pseudospin 1/2 system.

As follows from the Ehrenfest's theorem, for a system subjected to a potential $V(\{q_g\})$ with $\langle V(\{q_g\})\rangle =V(\langle \{q_g\}\rangle)$ (where $\langle  \cdots \rangle$ is the quantum mechanical  average   and $\{q_g\}$ stands for  generalized coordinates), the dynamics of a quantum observable follows its classical counterpart. However, on a time scale larger than the Ehrenfest time, classical nonlinear system and it's quantum counterpart manifest different features\cite{PhysRevE.65.035208}

We note that the nonlinear phenomenon plays an incisive role for NEMS.  Effects such as  Kerr-like nonlinearity \cite{Jacobs2009, Cleland2002} for mechanical resonators or the phononic nonlinear regime in strong external fields becomes relevant \cite{Rips2014, Weber14}.
Traditionally, in physics and mathematics,  classical nonlinear systems have been studied intensively as they show a wide range of interesting phenomena \cite{Strogatz2015}, that are expected to be reflected in the quantum behavior when such classical systems are coupled to quantum ones.
For instance, Drummond and Walls \cite{Drummond1980} showed that a nonlinear model of an optical cavity being driven by a continuous external field shows a bistable window in its semiclassical description.  This can be contrasted with an analogous quantum system in which bistable regime is not present.

In this paper we investigate a paradigmatic model of NEMS hybrid system: a nonlinear oscillator coupled to a spin 1/2 system. We show that
in spite of the nonlinearity in the system, the exact analytical solution is accessible. We introduce a scheme of periodic driving and study the combined effects of the driving and coupling between the spin and the nonlinear oscillator. The advantage of the NV centers is their relatively low decoherence rate. However, on the longer run, even a low  decoherence may lead to substantial effects.  Thus, we use a simple unitary evolution protocol which (even though not reducing entirely the  decoherence) leads  to the generation of coherence.

The manuscript is divided into the sections as follows:
In Section-\ref{sec:model}, we discuss the model and transformation of the
cantilever problem
to a mathematical pendulum. In Section-\ref{sec:MathieuS}, we formulate
the Mathieu Schr\"odinger equation. Next, in section-\ref{sec:analytical1}, we discuss
the spin dynamics of the NV center. Subsequently, in section \ref{sec:lindblad}, we study the effects of the environment with a Lindblad master equation and non-Markovian noise due to $^{13}C$ nuclei in the surroundings of NV spin.  Section-\ref{multilevel} is about multilevel dynamics and entanglement,  followed by Section-\ref{unitary_coherence} that addresses  the unitary generation of coherence. Section \ref{summary} summarizes.
\section{Theoretical modeling}
\label{sec:model}
The Hamiltonian of the NV center coupled to a driven nonlinear oscillator reads \cite{Zhou2010}
\begin{eqnarray}\label{Lorentz}
H\big(x,p,t\big)&=&H_{S}+H_{0}\big(x,p\big)+H_{NL}\nonumber \\
&+&\varepsilon V\big(x,t\big)+g \cos(\omega t)x S_{z}.
\end{eqnarray}
Here $H_{s}=\frac{1}{2}\omega_{0} \sigma_{z}$ is the Hamiltonian of the NV center, $\omega_{0}=\big(\omega_{R}^{2}+\delta^{2}\big)^{1/2}$, $\omega_{R}$ is the Rabi frequency,
and $\delta$ is the detuning between the microwave frequency and the intrinsic frequency of the spin. In what follows,we set $\hbar$ equal to one. The operator $S_{z}$ in the eigenbasis of the NV center has the form:
$S_{z}=\frac{1}{2}\big(\cos(\alpha)\sigma_{z}+\sin(\alpha)(\sigma_{+}+\sigma_{-})\big)$ with $\tan(\alpha)=-\omega_{R}/\delta$ and $\sigma_{z}=|e\rangle\langle e|-|g\rangle\langle g|$, $\sigma_{+}=|e\rangle\langle g|,~~\sigma_{-}=|g\rangle\langle e|$. For more details see \cite{Zhou2010}. The terms $H_{0}$ and $H_{NL}$ represent linear and nonlinear parts of the oscillator respectively:
\begin{equation}
H_{0}=\frac{p^{2}}{2m}+\frac{\omega_{r}^{2}mx^{2}}{2},~~~H_{NL}=\beta x^{3}+\mu x^{4},
\label{hamiltonian1}
\end{equation}
where $\omega_{r}$ is the frequency of the oscillations, $\beta$ and $\mu$ are constants of the nonlinear terms.
The term
\begin{equation}
V\big(x,t\big)=V_{0}x\cos\omega t,~~~\varepsilon V_{0}=f_{0},~~\varepsilon \ll 1.
\label{hamiltonian2}
\end{equation}
describes the driven motion of the cantilever in the RF field with frequency $\omega$.
The last term in Eq. (\ref{Lorentz}) describes the coupling between the oscillator and the NV spin. The distance and the coupling strength between the magnetic tip and NV spin can be modulated through the magnetostriction effect \cite{Chotorlishvili2013}.

Our description of the problem is quite general. However, without the loss of generality we specify the values of the parameters relevant for the NV centers\cite{Rabl2009}: $\frac{\omega_{r}}{2\pi}=5$ MHz, $\frac{\omega_{R}}{2\pi}=0.1-10$  MHz, $\delta=1$ kHz, mass of the cantilever $m=6\times 10^{-17}$ kg, the coupling constant $\frac{g}{2\pi}=100$ kHz, the amplitude of the zero point fluctuations $a_{0}=\sqrt{\hbar/2m\omega_{r}}\approx 5\times 10^{-3}$ m. The nonlinear constants are order of $\beta\approx\frac{\omega_{r}^{2}m}{2a_{0}}$, $\mu\approx\frac{\omega_{r}^{2}m}{2a_{0}^{2}}$. The energy scale of the problem is defined by
$\varepsilon V\approx \omega_{r}^{2}ma_{0}^{2}\approx 10^{-9}$J, and the time scale is of order of microsecond scale $t\approx \frac{\pi}{2g}$ microseconds.

\subsection{Classical cantilever dynamics}
Let us discuss the dynamics of classical nonlinear cantilever using the Hamiltonian given by Eqs. (\ref{hamiltonian1}) and (\ref{hamiltonian2}).  The equation of motion governed by $H_{0}+H_{NL}+V(x,t)$ has the form:
\begin{equation}\label{L1}
\ddot{x}+\omega_{r}^{2}(1+\alpha_{3} x+ \alpha_{4} x^{2})x=V_{0}\cos(\omega t).
\end{equation}
For brevity we introduced the notations $\alpha_{3}=3\beta/\omega_r^2, \quad \alpha_{4}=4\mu/\omega_r^2.$
Adopting the perturbation ansatz
 $$x(t)=x^{(1)}(t)+x^{(2)}(t)+x^{(3)}(t)+\cdots,~x^{(1)}(t)=V_{0}\cos(\Omega t),$$ where $$\Omega=\omega_{r}+\Omega_{1}+\Omega_{2}+\cdots,$$ the equation (\ref{L1}) now takes on the form:
\begin{eqnarray}\label{L2}
\big(\omega_{r}/\Omega\big)^{2}\ddot{x}+\omega_{r}^{2}x&=&-\alpha_{3}\omega_{r}^{2}x^{2}-
\alpha_{4}\omega_{r}^{2}x^{3}\nonumber\\
&-& \big(1-\omega_{r}^{2}/\Omega \big)\ddot{x}+V_{0}\cos(\omega t).
\end{eqnarray}
Suppose that $\omega=\omega_{r}/3+\Delta\Omega$, where $\Delta\Omega$ is the small modulation frequency in the vicinity of the resonance $n\Omega(x)\approx m \omega$, where $n,~m\in\mathbb{Z}$. Then the first order term $x^{(1)}(t)\sim \cos\big[3\big(\omega_{r}/3+\Delta\Omega\big)\big]$ is off-resonance. However, the second-order term $\{x^{(1)}(t)^{3}\}\sim\cos\big[3\big(\omega_{r}/3+\Delta\Omega\big)\big]$ already leads to the parametric resonance. For our convenience we switch to the canonical pair of  action-angle $(I, \theta)$ variables. The cantilever part of the Hamiltonian $H_{p,q}=H_{0}+H_{NL}+V(x,t)$ expressed in new variables $H_{I,\theta}$ is connected to the original Hamiltonian through the production function $\Phi=F+I\theta$ via the relation:
\begin{eqnarray}\label{L3}
&& d\Phi=pdq+\theta dI+\big(H_{I,\theta}-H_{p,q}\big)dt,
\end{eqnarray}
and the canonical set of equations in new variables are:
\begin{equation}\label{L4}
  \left.\begin{aligned}
  \frac{dI}{dt}&=-\frac{\partial H_{I,\theta}}{\partial\theta}=-\varepsilon \frac{\partial V(I,\theta, \lambda)}{\partial \theta}, \\
  \frac{d\theta}{dt}&=\frac{\partial H_{I,\theta}}{\partial I}=\Omega(I)+\varepsilon \frac{\partial V(I,\theta, \lambda)}{\partial I}.
\end{aligned}\right\} 
\end{equation}

Here we introduced the nonlinear frequency $\Omega(I)=\partial (H_{0}+H_{NL})/\partial I$ and $\dot{\lambda}=\omega$ the frequency of external driving. Nonlinearity of the system is quantified by the following criterion:
\begin{eqnarray}\label{L5}
 \mathcal{A}&=&\bigg|\frac{I}{\Omega (I)}\frac{d\Omega(I)}{dI}\bigg|\nonumber \\
&=&\bigg|I\bigg(\frac{\partial (H_{ 0}+H_{NL})}{\partial I}\bigg)^{-1}\frac{d}{dI}\bigg(\frac{\partial(H_{0}+H_{NL})}{\partial I}\bigg)\bigg|.
\end{eqnarray}
where,
\begin{equation}
H_{0}(I)=\omega_{r}I+H_{NL},~~~H_{NL}=3\pi\bigg(\frac{I}{m\omega_{r}}\bigg)^{2}\mu.
\label{2transformd}
\end{equation}

The deviation of action from the resonance value is given by $\Delta I=I-I_{0}$. The nonlinear frequency $\Omega(I)=\partial H_{0}/\partial I+\partial H_{NL}/\partial I$ and the nonlinear resonance condition in the action-angle variables has the form $\omega_{r}+\Omega_{NL}(I_{0})=\omega$, $\Omega_{NL}=6\pi I(\mu/m^{2}\omega_{r}^{2})$.

Our method is valid if $1/\varepsilon\gg\mathcal{A}\gg \varepsilon$. \cite{zaslavsky2007physics}
To explore the nonlinear multiple resonances, we utilize the standard expansion adopted in the theory of dynamical systems \cite{zaslavsky2007physics}
\begin{eqnarray}\label{L6}
 \varepsilon V(I,\theta, \lambda)=
\frac{1}{2}\sum\limits_{n,m}V_{n,m}(I)\exp\big[i(m\lambda+n\theta)\big]+c.c.\nonumber \\
\end{eqnarray}
and insert  Eq. (\ref{L6}) in Eq. (\ref{L5}) to find
\begin{equation}\label{L7}
\left. \begin{aligned}
\frac{dI}{dt}&=\varepsilon n V_{mn}\sin(\digamma_{mn}),\\
\frac{d\digamma_{mn}}{dt}&=m\omega+n\Omega(I)+\varepsilon n \bigg(\frac{\partial V_{mn}}{\partial I}\bigg)\cos(\digamma_{mn}).
\end{aligned}\right\}
\end{equation}
Here $\digamma_{mn}=m\lambda+n\theta$ is the resonant phase. Let $I_{0}$ correspond to the exact resonance $\digamma_{mn}=0,~|\triangle I-I_{0}|=|\triangle I|\ll I_{0}$.  The set of equations Eq. (\ref{L7}) for the deviation of the action $\triangle I$ becomes
\begin{equation}\label{L8}
\left. \begin{aligned}
\frac{d(\triangle I)}{dt}&=\varepsilon nV_{mn}\sin(\digamma_{mn}),\\
\frac{d(\digamma_{mn})}{dt}&=n\big[\Omega(I)-\Omega(I_{0})\big]\\\
&+\varepsilon n\bigg[\frac{\partial V_{mn}}{\partial I}\cos(\digamma_{mn})
-\frac{\partial V_{mn}(I_{0})}{\partial I}\bigg].
\end{aligned}\right\}
\end{equation}
Taking into account that $\Omega(I)-\Omega(I_{0})\approx \big(\frac{\partial \Omega}{\partial I}\big)_{I=I_{0}}\triangle I=\mathcal{A}\Omega(I_{0})\triangle I/I_{0}$ and the condition of the moderate nonlinearity $1/\varepsilon\gg\mathcal{A}\gg \varepsilon$, in Eq. (\ref{L8}) , we find:

\begin{equation}\label{L9}
\left. \begin{aligned}
\frac{d \triangle I}{dt}&=\varepsilon n V_{mn}\sin(\digamma_{mn}),\\
 \frac{d \digamma_{mn}}{dt}&=n\bigg(\frac{\partial \Omega (I)}{\partial I}\bigg)_{I=I_{0}}.
\end{aligned}\right\}
\end{equation}
These are the Hamilton's equations with the  Hamiltonian

\begin{eqnarray}\label{L10}
&& H_{I,\theta}=n\bigg(\frac{d\Omega (I)}{dI}\bigg)_{I_{0}}\frac{(\triangle I)^{2}}{2}+\varepsilon n V_{mn}(I_{0})\cos(\digamma_{mn}).\nonumber \\
\end{eqnarray}
For simplicity we use the rotating wave approximation and retain the slow phase in the oscillator-spin coupling term $\varphi=\theta-\omega t$.  Considering only the first resonance $n=m=1$ from Eq. (\ref{Lorentz})  we deduce the transformed total Hamiltonian as
\begin{equation}
H=H_{s}+H_{0}(I)+\big\{\varepsilon V(I)+Q(I)S_{z}\big\}\cos(\varphi).
\label{transformd}
\end{equation}
Here, for brevity,  the following notations are used
\begin{equation}
\varepsilon V(I)=V_{0}\sqrt{I/m\omega_{r}},~Q(I)=g \sqrt{I/m\omega_{r}}.
\label{3transformd}
\end{equation}
\section{Quantum cantilever dynamics}
The transformed Hamiltonian of Eq. (\ref{Lorentz}) can be written as
\begin{equation}\label{total1}
H=H_{m}+H_{s}+Q\cos\varphi S_{z},
\end{equation}
where $H_m$ is the Hamiltonian of a mathematical pendulum given by
\begin{equation}
H_{m}=\frac{\omega'}{2}\big(\triangle I\big)^{2}+U\cos\varphi,
\label{total}
\end{equation}
with the  notations $\omega'=\big(d\Omega_{NL}(I)/dI\big)\mid_{I=I_{0}}$, $U=\varepsilon V\big(I_{0}\big)$.
We are interested in the analytical solutions to the Hamiltonian Eq. (\ref{total}).
  The Schr\"odinger equation with the Hamiltonian $H_m$, is related to corresponding Mathieu-Schr\"odinger equation and its spectrum.
\begin{figure}[t!]
 \includegraphics[width=\linewidth]{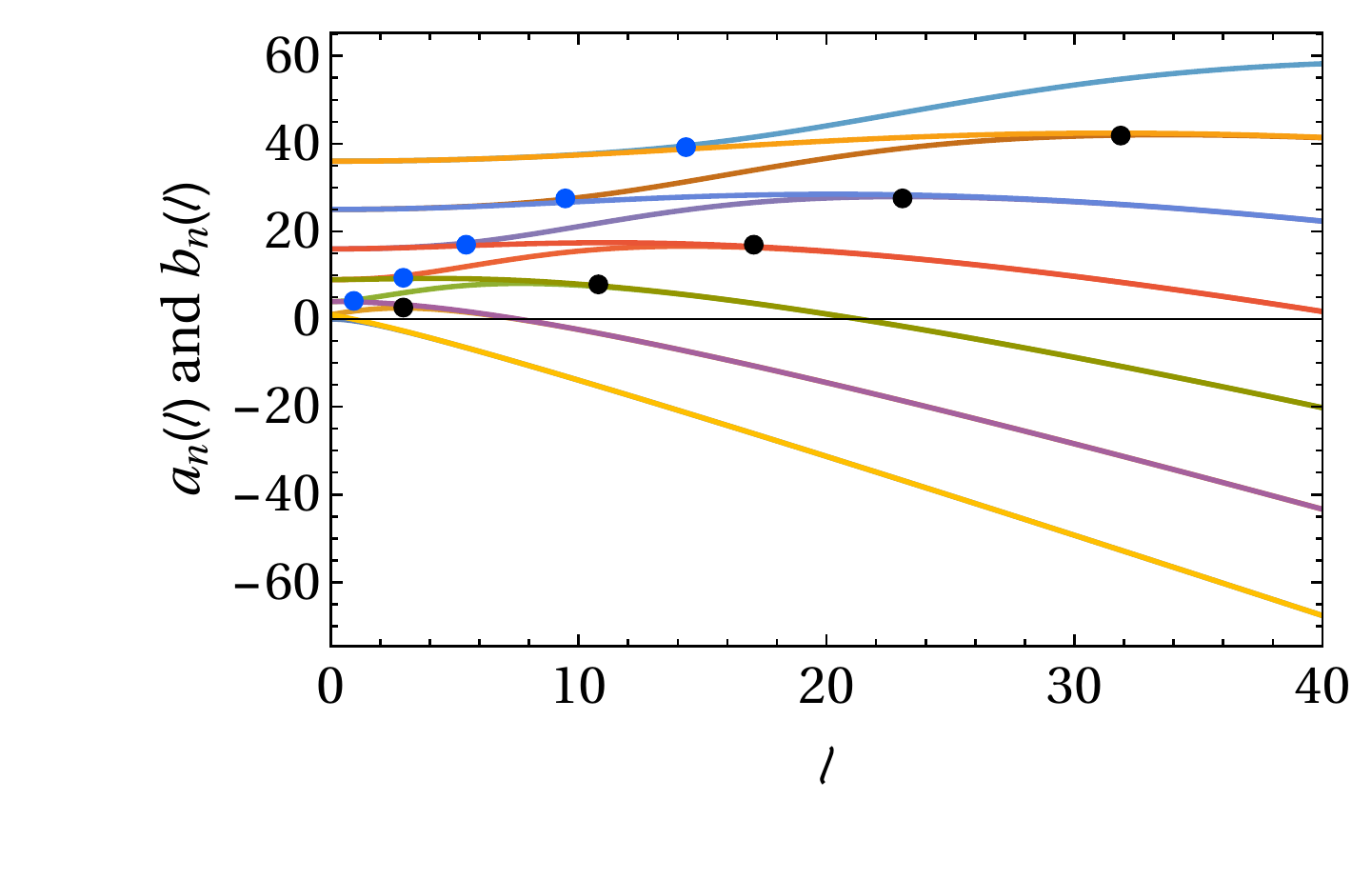}
\caption{Energy spectrum $E_{n}(l)$ of Mathieu-Schr\"odinger equation with varying  barrier height $l$. The region where curves are split   is called $G_{0}$ and the merging points define the boundaries of the $G_{-}$  and $G_{+}$ subgroups. The energy spectrum corresponding to Mathieu function $|ce_{n}(l,\varphi)\rangle$
is described by Mathieu characteristic $a_{n}(l)$, and the energy spectrum corresponding to Mathieu function $|se_{n}(l,\varphi)\rangle$ is described by Mathieu characteristic $b_{n}(l)$.}
\label{fig1}
\end{figure}
\label{sec:MathieuS}
This becomes evident when applying
correspondence principle,  by relating  the classical variables to the corresponding operators, meaning
$\Delta I \rightarrow \imath \hbar\partial/\partial\varphi$ in Eq. (\ref{total}) which relates   the Hamiltonian of the mathematical pendulum $H_{m}$ to  the Mathieu-Schr\"odinger equation
\begin{equation}
\frac{d^{2}\psi_{n}}{d\varphi^{2}}+\big(E_{n}-2l\cos2\varphi)\psi_{n}=0.
\label{Mathieuequation}
\end{equation}
The  effective potential is  $V(l,\varphi)=2l\cos2\varphi$, and rescaled energy, the potential barrier and  the angle are $E_{n}\rightarrow \frac{8E_{n}}{\omega'}$, $l\rightarrow \frac{8U}{\omega'}$, $\varphi\rightarrow 2\varphi$, respectively.

A detailed analysis of the Mathieu-Schr\"odinger equation (\ref{Mathieuequation}) was done in numerous  works, for example the references  \cite{Ugulava2005,Chotorlishvili2010,Chotorlishvili2018,horne1999}. The energy spectrum
of the Mathieu-Schr\"odinger equation  depends parametrically on the potential barrier $E_{n}\big(l\big)$ and contains two degenerate $G_{-},~~G_{+}$ and one non-degenerate domain $G_{0}$ (See Fig. \ref{fig1}).
The eigenfunctions of each region $G_{-},~~G_{+}$ and $G_{0}$ are the basis functions of the irreducible representations  of the invariant subgroups of  Klein's four-group:
\begin{eqnarray}
&&G_{0}\subset \mathcal{E},\mathcal{A}, \nonumber \\
&&G_{-}\subset \mathcal{E},\mathcal{C}, \nonumber \\
&&G_{+}\subset \mathcal{E},\mathcal{B}.
\label{Klein}
\end{eqnarray}
and the group elements are
\begin{eqnarray}
&&G\big(\varphi\rightarrow -\varphi\big)=\mathcal{A},~~~G\big(\varphi\rightarrow \pi-\varphi\big)=\mathcal{B}, \nonumber \\
&&G\big(\varphi\rightarrow \pi+\varphi\big)=\mathcal{C},~~~G\big(\varphi\rightarrow \varphi\big)=\mathcal{E},
\label{Mathieuequation2}
\end{eqnarray}
while the irreducible basis functions for each subgroup are
\begin{equation}\label{eigenG-}
G_{-}\rightarrow |\phi^\pm_n (\varphi,l)\rangle =\frac{1}{\sqrt{2}}\big(ce_n(l,\varphi) \pm i se_n(l,\varphi)\big),
\end{equation}

\begin{equation}\label{eigenG0}
G_{0}\rightarrow   ce_n(l,\varphi),  se_n(l,\varphi),
\end{equation}
and
\begin{equation}\label{eigenG+}
G_{+}\rightarrow |\psi^\pm_n (\varphi,l)\rangle =\frac{1}{\sqrt{2}}\big(ce_n(l,\varphi) \pm i se_{n+1}(l,\varphi)\big).
\end{equation}
Here, $ce_n(l,\varphi)$ and $se_n(l,\varphi)$ are Mathieu functions with characteristic values $a_n(l)$ and $b_n(l)$, respectively. The trigonometric series representation of Mathieu functions are given as \cite{bateman1953higher}:
\begin{equation}
ce_{2m}(l,\phi) = \sum_{r=0}^\infty A_{2r}^{(2m)}(l) \cos(2r\varphi),
\label{furier1}
\end{equation}
\begin{equation}
ce_{2m+1}(l,\phi) = \sum_{r=0}^\infty A_{2r+1}^{(2m+1)}(l) \cos((2r+1)\varphi),
\label{furier2}
\end{equation}
\begin{equation}
se_{2m+1}(l,\phi) = \sum_{r=0}^\infty B_{2r+1}^{(2m+1)}(l) \sin((2r+1)\varphi),
\label{furier3}
\end{equation}
\begin{equation}
se_{2m+2}(l,\phi) = \sum_{r=0}^\infty B_{2r+2}^{(2m+2)}(l) \sin((2r+2)\varphi).
\label{furier4}
\end{equation}
\section{Quantum spin dynamics of NV center}
\label{sec:analytical1}
The system of the nonlinear oscillator coupled with the NV center spin is transformed into a system of a mathematical pendulum coupled with the NV center spin. In the previous section we have found the eigenfunctions and eigenvalues of the mathematical pendulum in terms of the Mathieu functions and the characteristic values of the Mathieu-Schr\"odinger equation.  To  explore the total Hamiltonian $\hat{H}$, Eq.(\ref{total})  we  use the joint basis of eigenfunctions of the mathematical pendulum
Eq.(\ref{eigenG-})-Eq. (\ref{eigenG+}) and the  basis functions of  Pauli matrix $\sigma_z$ for the spin part {\it i.e.,} $|\chi\rangle\equiv \{ |1\rangle,~|0\rangle\}$.
The total Hamiltonian  Eq. (\ref{total1}) reads
\begin{equation}
\label{allG-ham}
\hat{H}  =   \begin{pmatrix}
   A_{11} & A_{12} & A_{13} & A_{14}  \\
   A_{21} & A_{22} & A_{23}& A_{24}  \\
   A_{31} & A_{32} & A_{33}& A_{34}   \\
   A_{41} & A_{42} & A_{43}& A_{44}   \\
   \end{pmatrix},
\end{equation}
where the matrix elements $A_{ij}$ for the region $G_-$, $G_0$ and $G_+$ are presented in the Appendix \ref{appendix1}.
\begin{figure*}[t!]
 \includegraphics[width=\linewidth,height=2.2in]{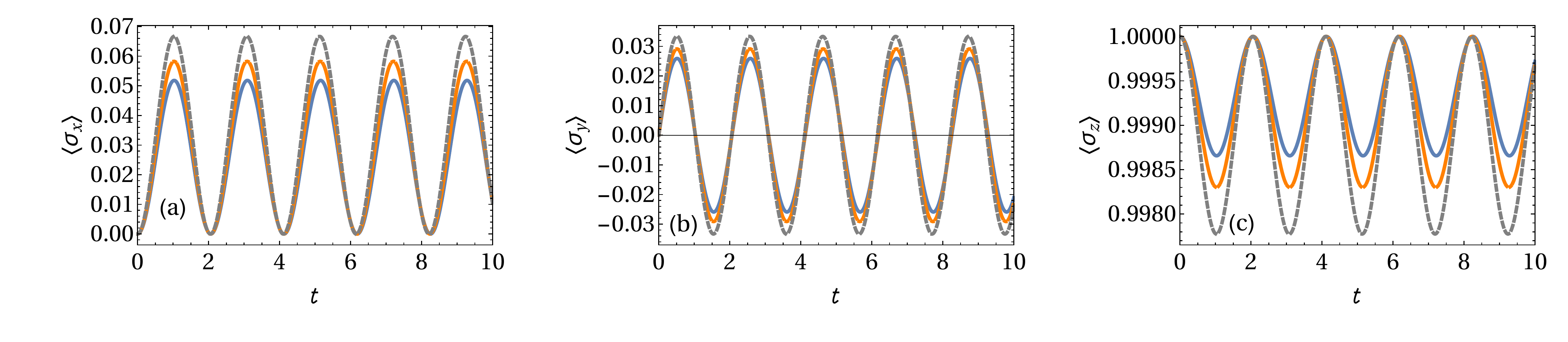}
\caption{ (a) Average transverse spin component $ \langle \sigma_{x} \rangle $,  (b) average transverse spin component $ \langle \sigma_{y} \rangle $, and (c) average longitudinal spin component $ \langle \sigma_{z} \rangle $, plotted for the bipartite system $\hat{\rho}_{AB}$ in the region $ G_{0} $ for different quantum numbers $n=2,3,4$. In all the figures blue (solid), orange (solid) and violet (dashed) lines represent $n=2,l=3.855$,  $n=3,l=7.535$ and  $n=4,l=10.785$ cases, respectively. The values of the barrier heights $l$ are chosen to be in $G_{0}$ region for the given $n$. The interaction strength between the nonlinear oscillator and the NV spin is taken to be $ Q = 0.5 $. Time is in the units of $\omega_0^{-1}$. }
\label{figure1}
 \end{figure*}
In the region $G_{0}$, for a given quantum number $n$, the system can be found either in the states $ \vert ce_{n}(\varphi,l) \rangle \otimes  \vert 0 \rangle  $ and $ \vert ce_{n}(\varphi,l) \rangle \otimes \vert 1  \rangle $, or  in the states $ \vert se_{n}(\varphi,l) \rangle \otimes  \vert 0 \rangle  $ and $ \vert se_{n}(\varphi,l) \rangle \otimes \vert 1  \rangle $.
We note that these states correspond to a particular fixed quantum number $n$. The presence of the spin-oscillator coupling term leads to the mixing of the nonlinear oscillator states. However, when the distance between the neighbor states is larger than the spin-oscillator coupling strength $E_{n+1}-E_{n}>g$, the states with a different quantum number are eliminated from the process.
Suppose that the system is in the states $ \vert ce_{n}(\varphi,l) \rangle \otimes  \vert 0 \rangle  $ and $ \vert ce_{n}(\varphi,l) \rangle \otimes \vert 1  \rangle $, then the Hamiltonian of the system in $G_0$ region can be written as:
\begin{equation}
\label{ham_G0}
\hat{H}=
\begin{pmatrix}
a_{n}(l)+\frac{1}{2}\omega_{0}+\frac{1}{2} Q e \cos {\alpha} &\frac{1}{2} Q e \sin {\alpha}  \\
\frac{1}{2}Q e \sin {\alpha} & a_{n}(l)-\frac{1}{2}\omega_{0}-\frac{1}{2} Q e \cos {\alpha}
\end{pmatrix}.\\
\end{equation}
The eigenvalues of the above Hamiltonian (\ref{ham_G0}) are $a \pm \sqrt{b^{2} + c^{2}}$ and the corresponding eigenvectors are $|\phi_{1}\rangle=|ce_{n}\rangle\left(\alpha_{1}|1\rangle+\beta_{1}|0\rangle\right)$,
$|\phi_{2}\rangle=|ce_{n}\rangle\left(\beta_{1}|1\rangle-\alpha_{1}|0\rangle\right)$,
 where $\alpha_{1}=1/\sqrt{\lambda^{2}+1}$, $\beta_{1}=\lambda/\sqrt{\lambda^{2}+1}$,
  $\lambda=(b+\sqrt{b^{2}+c^{2}})/c$, $a =  a_{n}(l) $ is the energy spectrum corresponding to Mathieu function $|ce_{n}(l,\varphi)\rangle$, $b  = \frac{1}{2} \omega_{0} + \frac{1}{2} Q e \cos {\alpha} $ , $ c =\frac{1}{2} Q e \sin {\alpha}$ and
\begin{equation}
e =\{\frac{\pi}{2}\left(A_{1}^{(2n+1)}(l)\right)^{2}+
 \pi\sum\limits_{r=0}^{\infty}A_{2r+1}^{(2n+1)}(l)A_{2r+3}^{(2n+1)}(l)\}.
\end{equation}
If the system is in the state $ \vert se_{n}(\varphi,l) \rangle \otimes  \vert 0 \rangle  $ and $ \vert se_{n}(\varphi,l) \rangle \otimes \vert 1  \rangle $, the Hamiltonian is
\begin{equation}
\hat{H}=
\begin{pmatrix}
b_{n}(l)+\frac{1}{2}\omega_{0}+\frac{1}{2} Q f \cos {\alpha} &\frac{1}{2} Q f \sin {\alpha}  \\
\frac{1}{2}Q f \sin {\alpha} & b_{n}(l)-\frac{1}{2}\omega_{0}-\frac{1}{2} Q f \cos {\alpha}
\end{pmatrix}.\\
\end{equation}
The eigenvalues and corresponding eigenvectors in this case are $a_{x} \pm \sqrt{[b_{x}^{2} + c_{x}^{2}]}$ and $ \Big[{\lambda_{x}/\sqrt{\lambda_{x}^2+1}, 1/\sqrt{\lambda_{x}^2+1}}\Big] $, $\lambda_{x}=(b_{x} \pm \sqrt{[b_{x}^2 + c_{x}^2]})/c_{x}$\\
Where $ a_{x} = b_{n}(l) $ is the energy spectrum corresponding to Mathieu function $|se_{n}(l,\varphi)\rangle$,
$ b_{x}= \frac{1}{2} \omega_{0} + \frac{1}{2} Q f \cos {\alpha} $, $ c_{x} =\frac{1}{2}Q f \sin {\alpha} $, and
\begin{equation}
f =\{\frac{\pi}{2}\left(-B_{1}^{(2n+1)}(l)\right)^{2}+
 \pi\sum\limits_{r=0}^{\infty}B_{2r+1}^{(2n+1)}(l)B_{2r+3}^{(2n+1)}(l)\}.
\end{equation}

Similarly in the degenerate $  G_- $ region the total Hamiltonian (\ref{allG-ham}) takes the form:
\begin{equation}
\label{ham_G-}
 \hat{H}=
\begin{pmatrix}
a_{-}+b_{1} & e_{1} & c_{1} & d_{1} \\
e_{1} & a_{-}-b_{1}  & d_{1} & -c_{1}\\
c_{1} &  d_{1} & a_{+}+b_{1} & e_{1} \\
d_{1} & -c_{1} & e_{1} & a_{+}-b_{1}
 \end{pmatrix},
\end{equation}
where $a_{\mp} = a_{1}= a_{n}(l)+b_{n}(l)$, $ b_{1}=\frac{1}{2} \omega_{0} + \frac{1}{4} Q r \cos {\alpha} $, $ c_{1}=\frac{1}{4} f_{1} Q \cos {\alpha} $, $ d_{1} =\frac{1}{4} f_{1} Q \sin {\alpha} $, $ e_{1} =\frac{1}{4} Q r \sin {\alpha} $,
\begin{eqnarray}\label{coefficients1}
r &=&\bigg[\{\frac{\pi}{2}\left(A_{1}^{(2n+1)}(l)\right)^{2}\nonumber\\&+&
 \pi\sum\limits_{r=0}^{\infty}A_{2r+1}^{(2n+1)}(l)A_{2r+3}^{(2n+1)}(l)\}+\{\frac{\pi}{2}\left(-B_{1}^{(2n+1)}(l)\right)^{2}\nonumber\\&+&
 \pi\sum\limits_{r=0}^{\infty}B_{2r+1}^{(2n+1)}(l)B_{2r+3}^{(2n+1)}(l)\} \bigg],
\end{eqnarray}
\begin{eqnarray}\label{coefficients2}
f_{1}&=&\bigg[\{\frac{\pi}{2}\left(A_{1}^{(2n+1)}(l)\right)^{2}\nonumber\\&+&
 \pi\sum\limits_{r=0}^{\infty}A_{2r+1}^{(2n+1)}(l)A_{2r+3}^{(2n+1)}(l)\}+\{\frac{\pi}{2}\left(-B_{1}^{(2n+1)}(l)\right)^{2}\nonumber\\&+&
 \pi\sum\limits_{r=0}^{\infty}B_{2r+1}^{(2n+1)}(l)B_{2r+3}^{(2n+1)}(l)\}  \bigg].
\end{eqnarray}
Four eigenvalues of the above Hamiltonian are
\begin{eqnarray}\label{coefficients3}
a_{1} \pm \sqrt{[b_{1}^2 - 2 b_{1} c_{1} + c_{1}^2 + d_{1}^2 - 2 d_{1} e_{1} + e_{1}^2]},
\end{eqnarray}
and
\begin{eqnarray}\label{coefficients4}
a_{1} \pm \sqrt{[b_{1}^2 + 2 b_{1} c_{1} + c_{1}^2 + d_{1}^2 + 2 d_{1} e_{1} + e_{1}^2]} .
\end{eqnarray}
The explicit form of the eigenvector corresponding to $G_{-}$ region is given  in  Appendix \ref{appendix4}.

Let us assume that the system is initially in the $G_{0}$ region and the state of the system is given by
\begin{eqnarray}
\vert \psi(0) \rangle = \vert ce_{n}(\varphi,l) \rangle\otimes \vert 0 \rangle.
\end{eqnarray}
The state of the system at any time $t$,  $\vert \psi(t) \rangle$ is the
solution of the Schr\"odinger equation using Hamiltonian Eq.(\ref{ham_G0}):
\begin{eqnarray}\label{1Schrodinger equation}
i \hbar \frac{d \vert \psi(t) \rangle}{dt} = \hat{H} \vert \psi(t) \rangle.
\end{eqnarray}
To solve Eq. (\ref{1Schrodinger equation}) one may  use the ansatz
\begin{equation}
\vert \psi(t) \rangle = C_{1}(t)\vert ce_{n} (\varphi,l)  \rangle \vert 0 \rangle + C_{2}(t) \vert ce_{n} (\varphi,l)  \rangle \vert 1 \rangle.
\label{2Schrodinger equation}
\end{equation}
\begin{figure*}
 \includegraphics[width=\linewidth,height=2.2in]{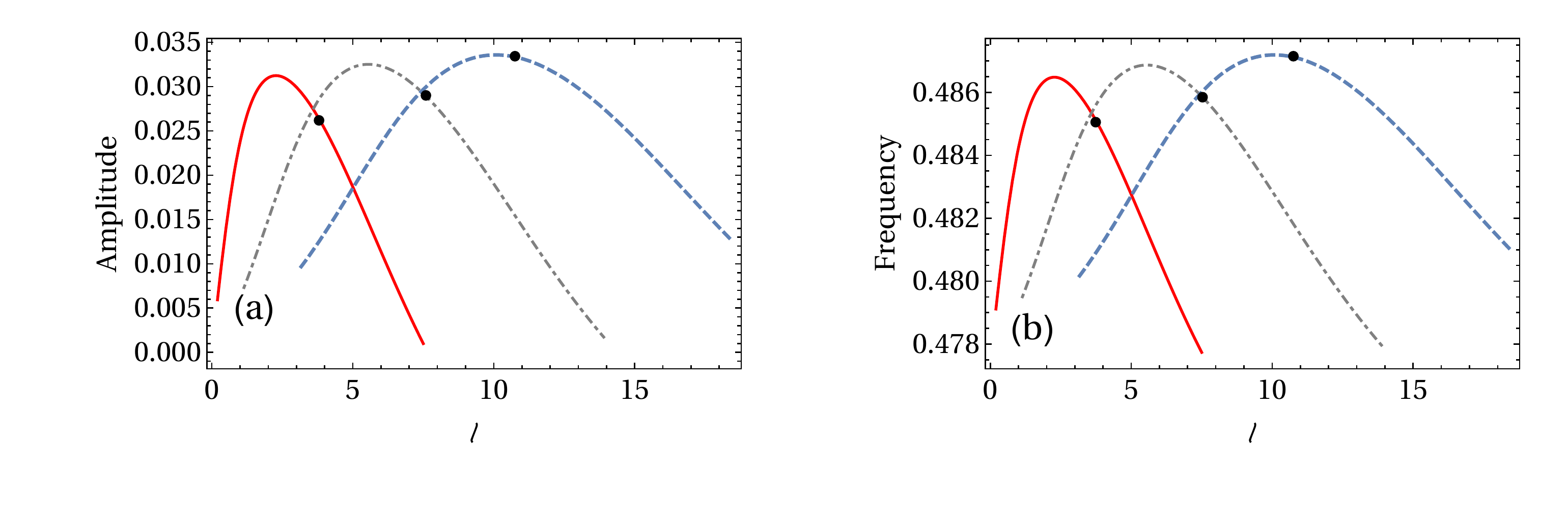}
\caption{(a) The amplitude and (b) frequency of $\langle \sigma_{x} \rangle $ with respect to barrier height  $l$ for different values of $n$. The quantum numbers $n$ and $l$ are chosen such that the mathematical pendulum remains in $G_{0}$ region. In all the figures, red (solid), violet (Dot-Dashed) and blue (dashed) lines represent $n=2$,  $n=3$ and  $n=4$ cases, respectively. The dots in the figures represent the points which are considered in Fig. \ref{figure1}.}
\label{figurexx}
 \end{figure*}
The  coefficients $C_{1}(t),~C_{2}(t)$ follow from Eqs.(\ref{1Schrodinger equation}), (\ref{2Schrodinger equation}).
The density matrix at any time $t$ can be calculated as $\rho_{AB}(t)=\vert \psi(t) \rangle\langle\psi(t)\vert$,
 $A$ and $B$ are used to emphasize the bipartite character of the system consisting of the
pendulum (A) and the spin parts (B).
We trace out the subsystem of the mathematical pendulum $\rho_{B}=Tr_{A}\rho_{AB}(t)$ and
the reduced density matrix is only the spin subsystem which is presented in a matrix form  as  
\begin{equation}
 \rho_{B}(t)=
\begin{pmatrix}
\rho_{11} & \rho_{12} \\ \rho_{21} & \rho_{22}
\end{pmatrix}.
\end{equation}
The elements of the density matrix are given by:
\begin{eqnarray}
\rho_{11}= \frac{b^2+c^2(\cos{\sqrt{b^2+c^2}t})^2} {(b^2+c^2)},
\end{eqnarray}
\begin{eqnarray}
&&\rho_{12}=\rho_{21}^*\\\nonumber&&=\frac {  c (b-b\cos{2 \sqrt{b^2+c^2}t}-i \sqrt{b^2+c^2}\sin{2 \sqrt{b^2+c^2}t})} {2(b^2+c^2)},
 \end{eqnarray}
 and
\begin{eqnarray}
 \rho_{22}= \frac{c^2(\sin^2{\sqrt{b^2+c^2}t})} {(b^2+c^2)}.
 \end{eqnarray}
Obviously, the state described by Eq. (\ref{2Schrodinger equation}) is a pure state.
Therefore, the purity that is defined as $\mathcal{P} = Tr(\rho_{B}^{2})=1$ and quantifies the mixedness
between the pendulum and the spin subsystem is equal to one.
We explore the dynamics of the expectation components of the spin $\langle\sigma_{\alpha} \rangle = Tr(\hat{\rho}\sigma_{\alpha})$, $\alpha=x,y,z$  and deduce
\begin{eqnarray}
 \langle \sigma_{z} \rangle_{G_{0}} = \frac{b^2+c^2\cos{2 \sqrt{b^2+c^2}{t}}}{b^2+c^2},
\end{eqnarray}
\begin{eqnarray}
 \langle \sigma_{y} \rangle_{G_{0}} = \frac{c\sin{2 \sqrt{b^2+c^2}{t}}}{\sqrt{b^2+c^2}},
\end{eqnarray}
\begin{eqnarray}
 \langle \sigma_{x} \rangle_{G_{0}} = \frac{2 b c\sin^2{ \sqrt{b^2+c^2}{t}}}{b^2+c^2}.
\end{eqnarray}
 In
Fig. \ref{figure1} (a), (b) and (c)  we show time evolution of $\langle \sigma_{x} \rangle_{G_{0}}$, $\langle \sigma_{y} \rangle_{G_{0}}$ and $\langle \sigma_{z} \rangle_{G_{0}}$ for different quantum states $n=2$, $3$ and $4$. The quantum number $l$ which  characterizes the barrier height is chosen carefully so that the system is near the separatrix line defined as $E_n=l$ in the $G_{0}$ region for given value of $n$.
From the above expressions of $\langle \sigma_{x} \rangle_{G_{0}}$, $\langle \sigma_{y} \rangle_{G_{0}}$ and $\langle \sigma_{z} \rangle_{G_{0}}$ it is evident that the modulation depth depends on $b$ and $c$ which varies with quantum numbers $n$ and $l$ through $e$. When the system is close to the separatix line the amplitude of the oscillations of transverse and longitudinal components of magnetization increases with increasing $n$ and the frequency of oscillation remains constant. Let us include even those points of $G_0$ region which are away from the separatrix line  corresponding to the given $n$. The amplitude of oscillations, for instance, for $\langle\sigma_x\rangle_{G_0}$ is given by $\frac{bc}{b^2+c^2}$, which follows a bell-shaped pattern as shown in Fig. \ref{figurexx} (a) for different values of $n$. The frequency of oscillation shows a similar behaviour as displayed in Fig. \ref{figurexx} (b).  We can observe a similar trend for $\langle\sigma_y\rangle_{G_0}$ and $\langle\sigma_z\rangle_{G_0}$ cases.

The dynamics of the system in the subgroup $G_{-}$ is much more complex.
Let us consider the system initially  in $G_{-}$ region and the state of the system is given as
 \begin{eqnarray}\label{x}
 \vert \psi(0) \rangle = \vert \phi_{n}^{-}(\varphi,l) \rangle \otimes \vert 0 \rangle.
 \end{eqnarray}
The state of the system at any time $t$ can be obtained using the  Eq. (\ref{1Schrodinger equation}). We consider the following ansatz for the state of the system in the region $G_{-}$:
\begin{eqnarray}
\vert \psi(t) \rangle&=&\zeta_{1}(t)\vert \phi^{-}_n (\varphi,l)  \rangle \vert 0 \rangle +\zeta_{2}(t)\vert \phi^{-}_n (\varphi,l)  \rangle \vert 1 \rangle \nonumber \\
&+&\zeta_{3}(t)\vert \phi^{+}_n (\varphi,l)  \rangle \vert 0 \rangle +\zeta_{4}(t)\vert \phi^{+}_n (\varphi,l)  \rangle \vert 1 \rangle.
\end{eqnarray}
and calculate the time dependent coefficients $\zeta_1(t)$, $\zeta_2(t)$, $\zeta_3(t)$ and $\zeta_4(t)$ satisfying the Schr\"odinger equation.   Now, we can calculate the density matrix $ \rho_{AB}(t)=\vert \psi(t) \rangle\langle \psi(t)\vert $ as
\begin{equation}
\rho_{AB}(t) =
\begin{pmatrix}
\rho_{11} & \rho_{12} & \rho_{13} & \rho_{14} \\
\rho_{21} & \rho_{22}  & \rho_{23} & \rho_{24}\\
\rho_{31} &  \rho_{32} & \rho_{33} & \rho_{34} \\
\rho_{41} & \rho_{42} & \rho_{43} & \rho_{44}
 \end{pmatrix}.
\end{equation}
Here the matrix elements of $\rho_{AB}(t)$ are constructed through the coefficients $\rho_{nm}=\zeta_{n}\zeta^{\ast}_{m}$.
The explicit expressions  for the coefficients $\zeta_{n}$ are obtained by solving the Schr\"odinger equation and separating the equations for the coefficients  is given in Appendix \ref{appendix5} .

\begin{figure*}[t!]
 \includegraphics[width=\linewidth,height=2.2in]{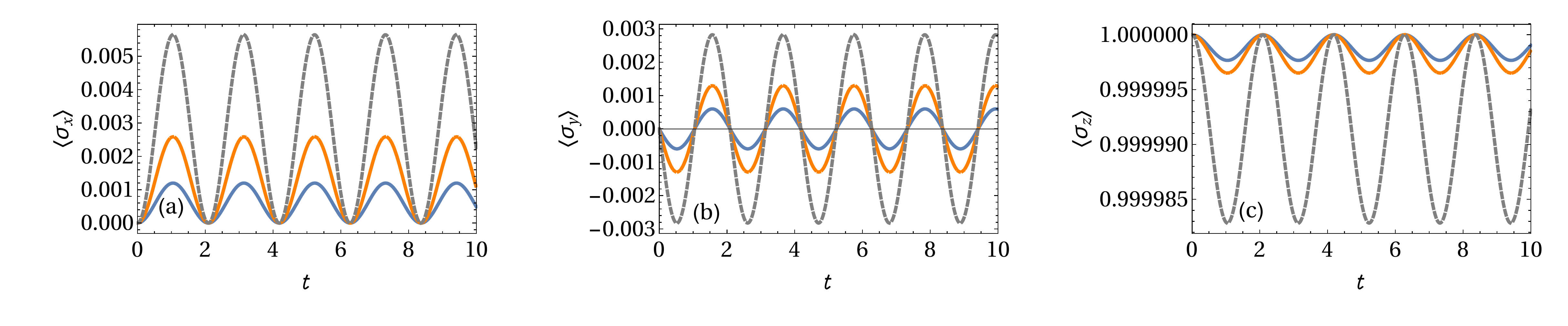}
\caption{(a) Average transverse spin component $ \langle \sigma_{x} \rangle $,  (b) average transverse spin component $ \langle \sigma_{y} \rangle $, and (c) average longitudinal spin component $ \langle \sigma_{z} \rangle $, plotted as a function of time for the bipartite system $\hat{\rho}_{AB}$ in the region $ G_{-} $ for different quantum numbers $n=2,3,4$. In all the figures blue (solid), orange (solid) and violet (dashed) lines represent $n=2,l=0.1$,  $n=3,l=0.57$ and  $n=4,l=1.585$ cases, respectively. In all the cases the coupling constant $Q$ is equal to $ 0.5 $. The values of barrier heights $l$ are chosen to be in the region $G_{-}$ for the given $n$. Time is in the units of $\omega_0^{-1}$.}
\label{figure5}
 \end{figure*}

For calculating the expectation values involving the spin part of the system we trace out the mathematical pendulum part $A$ in the region $G_{-}$. The reduced density matrix of the system is defined as $ \hat{\rho}_{B}(t)=Tr_{A}(\hat{\rho})$ and given by
\begin{eqnarray}
 \hat{\rho}_{B}(t)&=&\big(|\zeta_{1}|^{2}+|\zeta_{3}|^{2}\big)|0\rangle\langle0|+\big(|\zeta_{2}|^{2}+|\zeta_{4}|^{2}\big)|1\rangle\langle1|\nonumber\\
 &+&\big(\zeta_{1}\zeta_{2}^{\ast}+\zeta_{3}\zeta_{4}^{\ast}\big)|0\rangle\langle1|+h.c..
\end{eqnarray}
Let us calculate the expectation values of spin in the longitudinal and transverse directions defined earlier as $\langle\sigma_{\alpha} \rangle = Tr(\hat{\rho}_B\sigma_{\alpha})$, $\alpha=x,y,z$.
\begin{eqnarray}
 \langle \sigma_{z} \rangle_{G_{-}} &=&  |\zeta_1(t)|^2+|\zeta_3(t)|^2
 -|\zeta_2(t)|^2-|\zeta_4(t)|^2,
\end{eqnarray}
\begin{eqnarray}
\langle \sigma_{y} \rangle_{G_{-}} &=&-2Im(\zeta_1(t)\zeta_2^{*}(t)+\zeta_3(t)\zeta_4^{*}(t)),
\end{eqnarray}
\begin{eqnarray}
\langle\sigma_{x} \rangle_{G_{-}} &=&2Re(\zeta_1(t)\zeta_2^{*}(t)+\zeta_3(t)\zeta_4^{*}(t)).
\end{eqnarray}

Fig. \ref{figure5}  indicates that the spin dynamics in the region $G_{-}$ is similar
to the spin dynamics in the region $G_{0}$. In the region $G_{0}$ both the longitudinal and the transverse spin components show larger amplitude of oscillation in the excited states, while in the region $G_{-}$ the oscillation amplitudes of the longitudinal component $ \langle \sigma_{z}(t) \rangle $ are smaller than in the $G_{0}$ region (see Fig. \ref{figure5}(c)).
We see that $c_1$, $d_1$ and $e_1$ are negligibly small in comparison to $b_1$, therefore $\lambda_1^2\approx\lambda_2^2\approx b_1^2$. Taking this approximation into account, we can write
$\langle\sigma_{x} \rangle_{G_{-}}$ in a simpler form as
\begin{eqnarray}
\langle\sigma_{x} \rangle_{G_{-}}\approx 2(b_1e_1+c_1d_1)\frac{\sin^2\big(t\lambda_2\big)}{\lambda_2^2,},
\end{eqnarray}
where the amplitude of oscillation is $\frac{2(b_1e_1+c_1d_1)}{\lambda_2^2}$ and frequency of oscillation is $\frac{\lambda_2}{\pi}$. In this region, there is a small variation of the amplitude and the frequency of oscillation with the quantum numbers $n$ and $l$.  Since $G_-$ region is away from the line of separatrix, for a given $n$, the amplitude of oscillations are linearly increasing (though small)  and the frequency of oscillations are nearly constant with $l$. A similar description holds true for  $ \langle \sigma_{y}(t) \rangle_{G_-} $
and  $ \langle \sigma_{z}(t) \rangle_{G_-} $ cases.
\section{Dissipation}
\label{sec:lindblad}
In this section, we explore the decoherence processes. To keep a general discussion, we consider two different cases:
First, the relaxation process will be described by the Markovian master equation. However, we note that the relaxation processes described by the Markovian master equation are not the only source of decoherence in NV centers. The other case which is the primary cause of decoherence for the NV centers usually is due the nuclear-spin bath surrounding the electron spin (see \cite{de2010universal, cai2012robust} and references therein). At first, we consider the Markovian master equation that allows us to obtain an analytical result.

\subsection{Markovian Lindblad master equation}
To explore the decoherence  due to the environment the Lindblad master equation approach is used.
The Hamiltonian of the system is given by Eq. (\ref{total1}).
Let us suppose the system time evolution is nonunitary but Markovian. The nonunitary evolution of the system may cause  dissipation as the information may transmitted to the environment. The Liouville-von Neumann -Lindblad equation containing dissipation and decoherence terms  describe this nonunitary Markovian evolution of the system. One can start with Liouville- von Neumann-Lindblad master equation for the density matrix \cite{breuer2002theory,Mishra2014}
\begin{eqnarray}
\frac {d\rho}{dt} = -i[H  ,\rho]+
 \gamma (\sigma_{-} \rho \sigma_{+} - \frac{1}{2} (\sigma_{+} \sigma_{-} \rho + \rho \sigma_{+} \sigma_{-} )),\nonumber \\
\label{lindblad_eqn}
\end{eqnarray}
where $\gamma $ is  a dephasing parameter , and $ \sigma_{\pm}=\sigma_{x}\pm i \sigma_{y}$ are dissipators. Let us consider the system in $G_0$ region. Then the differential equations for each element of the density matrix can be obtained from the Eq. (\ref{lindblad_eqn}) as
\begin{eqnarray}
\dot{\rho}_{11} &=&-\gamma  \rho_{11} + i c (\rho_{12} - \rho_{21}),\nonumber\\
\dot{\rho}_{12} &=& -\frac{\gamma}{2}  \rho_{12} + i (-2 b \rho_{12} + c (\rho_{11} - \rho_{22})), \nonumber \\
\dot{\rho_{21}} &=& -\frac{\gamma}{2}  \rho_{21} + i (2 b \rho_{21} + c (\rho_{22} - \rho_{11})), \nonumber \\
\dot{\rho}_{22} &=&\gamma  \rho_{11} - i  c (\rho_{12} - \rho_{21}).
\end{eqnarray}
\begin{figure*}[t!]
 \includegraphics[width=\linewidth,height=2.2in]{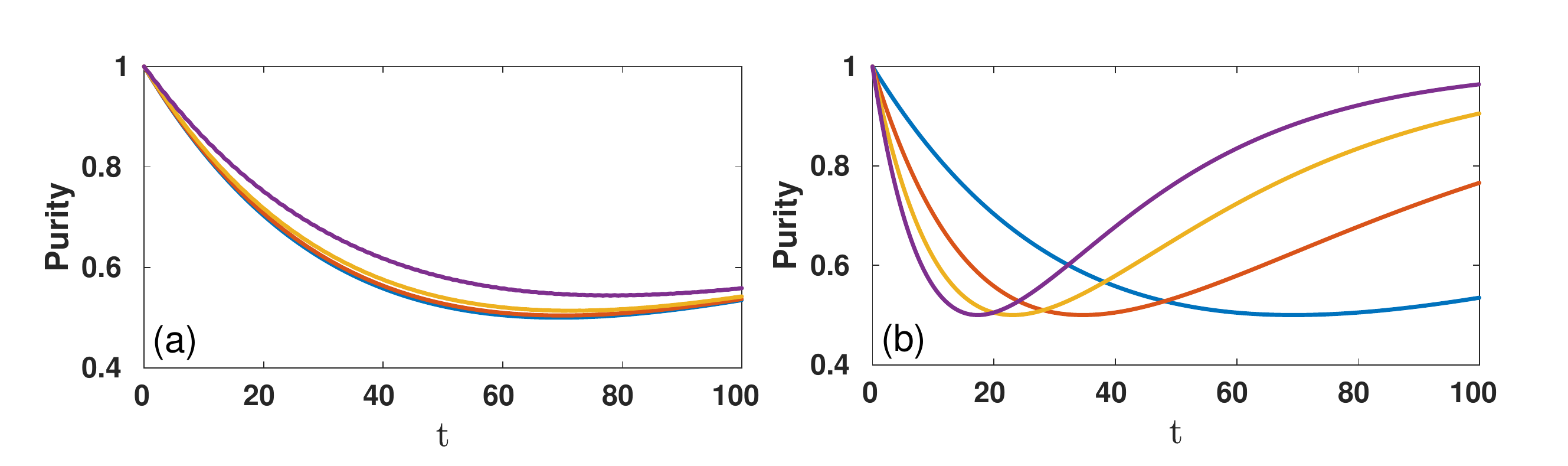}
\caption{Purity for the  hybrid system of NV center and nonlinear oscillator.
	(a) Behavior of purity for a damping constant $ \gamma = 0.01 $ at an arbitrary interaction strength  $Q$. The blue (solid), red (solid), yellow (solid) and purple (solid) lines represent $Q=0.5$,  $Q=5$, $Q=10$ and  $Q=25$ cases, respectively.
(b) For an interaction strength $Q=0.5$ but at arbitrary damping constant $\gamma$. The blue (solid), red (solid), yellow (solid) and purple (solid) lines represent  $\gamma=0.01$, $\gamma=0.02$, $\gamma=0.03$  and $\gamma=0.04$ cases, respectively. In  $ G_{0} $ region for the  quantum state $n=4$ . The barrier height  $l = 10.785$ corresponds to the region $ G_{0} $ in the vicinity of the transition into the region $G_{-}$. Time is in the units of $\omega_0^{-1}$.}
\label{figurex}
 \end{figure*}
 \begin{figure}[h]
  \includegraphics[width=\linewidth,height=2.2in]{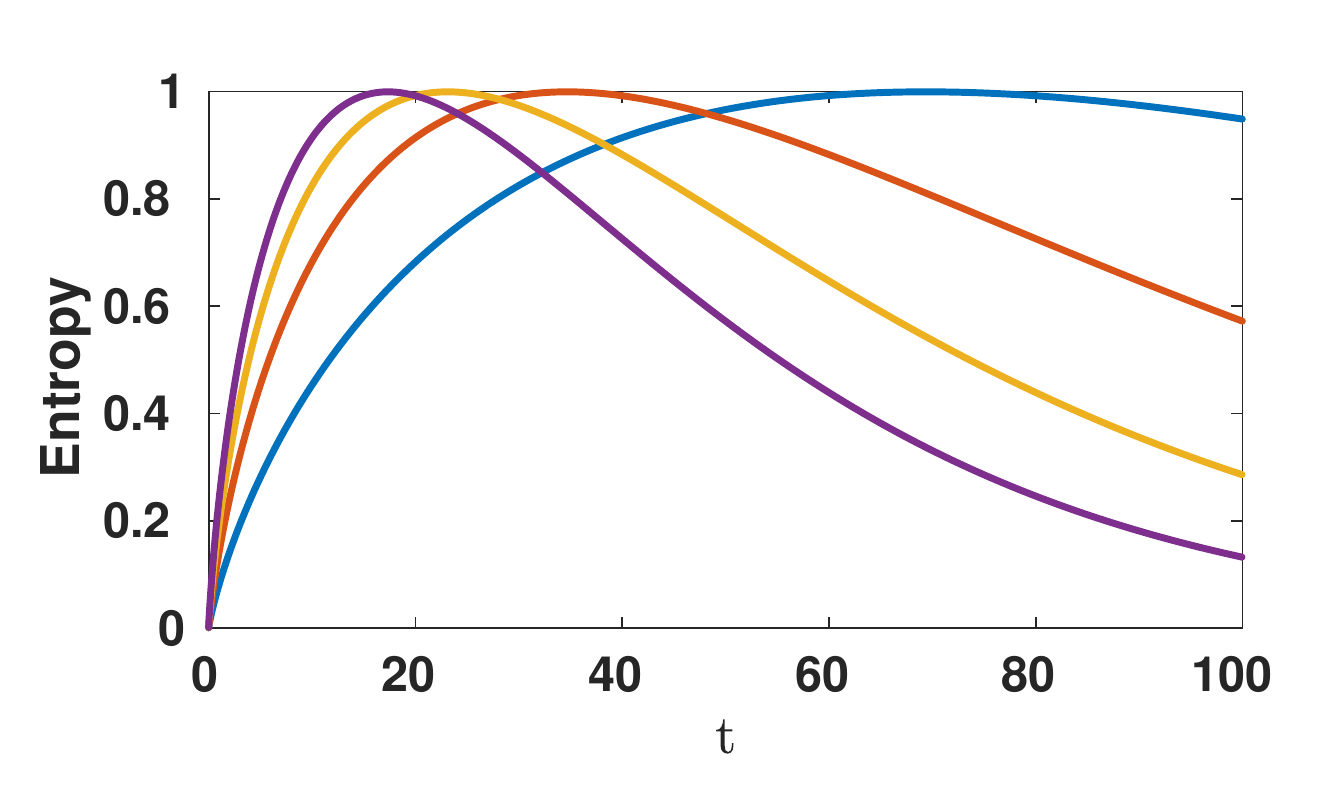}
\caption{ Entropy of the hybrid system of  NV center and nonlinear oscillator in the $ G_{0} $ region for the quantum state $n=4$. The barrier height  $l = 10.785$ corresponds to the region $ G_{0} $. The coupling strength $Q=0.5$ at arbitrary damping constant $\gamma$. The blue (solid), red (solid), yellow (solid) and purple (solid) lines represent  $\gamma=0.01$, $\gamma=0.02$, $\gamma=0.03$  and $\gamma=0.04$ cases, respectively. Time is in the units of $\omega_0^{-1}$.}
\label{entropy_t}
 \end{figure}
The solution to the above four equations can be obtained by using the corresponding initial conditions $\rho_{11}(0)=1$, $\rho_{12}(0)=0$, $\rho_{21}(0)=0$, $\rho_{22}(0)=0$ has the form
\begin{eqnarray}
\label{rho11t}
\rho_{11}(t)= \frac{b^2+c^2(\cos{\sqrt{b^2+c^2}t})^2} {(b^2+c^2)}e^{-\gamma t},
\end{eqnarray}

\begin{eqnarray}
\label{rho12t}
&&\rho_{12}(t)=\rho_{21}^*(t)\\\nonumber&&=\frac { c (b-b\cos{2 \sqrt{b^2+c^2}t}-i \sqrt{b^2+c^2}\sin{2 \sqrt{b^2+c^2}t})} {2(b^2+c^2)}e^{-\frac{\gamma}{2}t},
 \end{eqnarray}

\begin{eqnarray}
\label{rho22t}
\rho_{22}(t)= 1+(\frac{c^2(\sin{\sqrt{b^2+c^2}t})^{2}} {(b^2+c^2)}-1)e^{-\gamma t}.
 \end{eqnarray}
One can define the purity of  the NV spin coupled to the  mathematical pendulum and in contact with the environment as
 $\mathcal{P}(t,\rho)= Tr(\rho^2(t))$.
$\mathcal{P}(t,\rho)$ is a quantifier of mixedness of the system. Using the expressions for the density matrix elements evolved in time in accordance with Eqs. (\ref{rho11t})-(\ref{rho22t}), one can calculate $\mathcal{P}(t,\rho)$ as
\begin{widetext}
\begin{eqnarray}
\mathcal{P}(t,\rho)=\bigg(\frac{ (8 b^2 {\kappa}^2  +
   3 c^4 - (8 b^2 {\kappa}^2 + 3 c^4) e^{ \gamma t } +
   4 {\kappa}^4 e^{2 \gamma t } +
   c^2 (-1 + e^{\gamma t }) (-4 ( b^2 + {\kappa}^2 ) \cos { 2 \kappa t} -
      c^2 \cos{4 \kappa t}))} {4 {\kappa}^4}\bigg)e^{-2 \gamma t },\nonumber \\
\end{eqnarray}
\end{widetext}
where ${\kappa}^2 = b^2+c^2 $, $b$ and $c$ are defined in  Eq. (\ref{ham_G0}).
It is obvious that the isolated system ($\gamma=0$) for any arbitrary coupling strength $Q$ is always in a pure state.  However, for nonzero $\gamma$, one can observe that for any arbitrary coupling strength  $Q$, the system loses the purity as time goes on and evolves through the intermediate mixed state. Increasing the coupling strength enhances the purity, as seen in  Fig. \ref{figurex}(a).
\begin{figure*}[t]
  \includegraphics[width=\linewidth,height=2.2in]{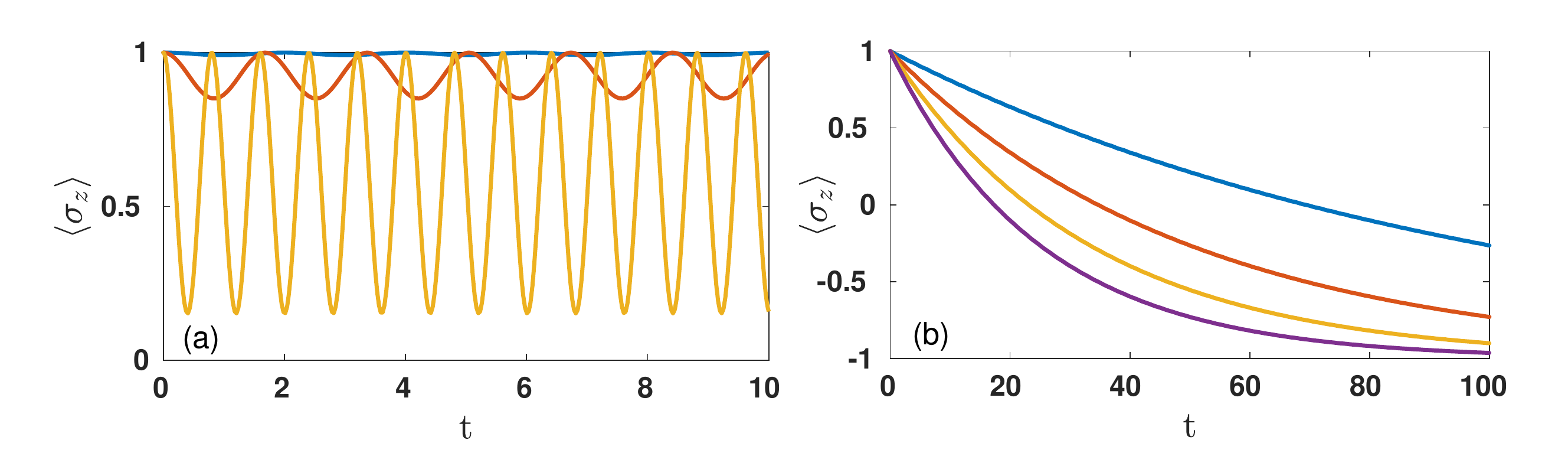}
\caption{Longitudinal spin component $\langle\sigma_{z}\rangle$ of the hybrid system of  NV center and nonlinear oscillator (a). The blue (solid), red (solid) and yellow (solid) lines represent  $Q=1$, $Q=5$ and $Q=25$ cases, respectively. For different damping coefficient $ \gamma$ and fixed coupling strength $ Q =0.5$ (b). The blue (solid), red (solid), yellow (solid) and purple (solid) lines represent  $\gamma=0.01$, $\gamma=0.02$, $\gamma=0.03$  and $\gamma=0.04$ cases, respectively.  Both  cases are in $ G_{0} $ region for the quantum state $n=4$ is considered. The barrier height  $l = 10.785$ corresponds to the region $ G_{0} $ in the vicinity of the transition to the $G_{-}$ region. Time is in the units of $\omega_0^{-1}$.}
\label{figurey}
 \end{figure*}

 Fig. \ref{figurex} (b) shows that the open quantum system for arbitrary $\gamma$ initially prepared in the pure state evolves through the intermediate mixed state, dips down to $0.5$ and finally reaches its value for the pure state. Interestingly, increasing the damping constants lowers the revival time of initial state.  The quantum revival of the state is also reflected in the dynamics of von Neumann entropy of the system (see Fig. \ref{entropy_t}). For the pure state the von Neumann entropy $S=-{\rm Tr} (\rho \log_2\rho) $ is zero, and for the completely mixed state, $S$ is one. Fig.\ref{figurex} (b) and Fig. \ref{entropy_t} show that for smaller damping coefficient, the system relaxes slowly and with more time needed for revival.

One can also analyze the dynamics of the longitudinal spin component $\langle\sigma_{z}\rangle$ as it is shown in Figs. \ref{figurey}(a) and \ref{figurey}(b).   Fig. \ref{figurey}(a)  shows that  the amplitude of the oscillation increases with increasing the coupling constant $Q$. Also, faster switching of the longitudinal spin component can be seen while increasing $Q$.
For fixed  coupling constant  $\langle\sigma_z\rangle$  does not oscillates but decays. Increasing the damping coefficient increases the decay rate (see Fig. \ref{figurey}(b)).

As we see, the master equation has a single steady state which is a pure state. Therefore, the dynamics asymptotically converges to a pure state.

\subsection{Fluctuations due to the spin bath}
The hyperfine coupling of the NV spin to the $^{13}$C nuclear spins causes a dephasing of the NV spin. This effect can be described by considering N independent reservoirs coupled to the NV spin.  It has been shown that the dynamics of the spin in the presence of the N reservoirs can be Markovian or non-Markovian\cite{breuer2002theory,PhysRevLett.105.050403,PhysRevLett.103.210401} depending upon the the number of reservoirs. If the number of reservoirs are above the cut off $N_c$, then the reservoirs act as a non-Markovian channel. $N_c$ depends on the bath parameters and the coupling between NV spin and reservoirs.
Let us consider the system in $G_{0}$ region is coupled to N independent bosonic reservoirs of field modes initially in the vacuum.\cite{PhysRevA.81.062124,PhysRevA.90.062104,PhysRevA.96.012105} The Hamiltonian of the system and the reservoir is given as
\begin{eqnarray}
\hat{H}=\hat{H}_{eff}+\sum_{n=1}^{N}\sum_{k}[\omega_{n,k}\hat{a}_{n,k}^{\dagger}\hat{a}_{n,k}\nonumber\\+g_{n,k}(\hat{a}_{n,k}^{\dagger}\hat{\sigma}^{-}+\hat{a}_{n,k}\hat{\sigma}^{+}],
\label{Hamiltonian_reservoir}
\end{eqnarray}
where $\hat{H}_{eff}=a\mathcal{I}-\sqrt{b^2+c^2} \hat{\sigma}^{z}$, $\mathcal{I}$ is the identity matrix, and the coefficients $a$, $b$ and $c$ in $\hat{H}_{eff}$, are given in Eq.(\ref{ham_G0}). In what follows we neglect the constant term and retain only the part involved in the spin dynamics $\hat{H}_{eff}=-\omega_{eff}\hat{\sigma}^{z}$, where $\omega_{eff}=\sqrt{b^2+c^2}$. The operator
$\hat{a}_{n,k}~(\hat{a}_{n,k}^{\dagger}) $ in Eq.(\ref{Hamiltonian_reservoir}) is annihilation (creation) operator of the $k^{\rm th}$ bosonic field mode of the $n^{\rm th}$ reservoir.

Let us consider the initial state of the system in a general form
\begin{eqnarray}
\vert \psi(0)\rangle=c_{0}(0)\vert0\rangle+c_{1}(0)\vert1\rangle,
\end{eqnarray}
where $c_0(0)$ and $c_1(0)$ are coefficients at $t=0$.  The state of reservoirs
is of the form $\prod_{n=1}^{N}\vert\Bar{0}\rangle_{n,r}$ with $\vert\Bar{0}\rangle_{n,r}=\prod_{k=1}{\vert0_{k}\rangle}_{n,r}$. Then the joint
state of the system and the reservoir is given by $\vert\Psi(0)\rangle=\vert \psi(0)\rangle\otimes\prod_{n=1}^{N}\vert\Bar{0}\rangle_{n,r}$ which after evolution becomes
\begin{eqnarray}
\vert\Psi(t)\rangle&=&[c_{0}(0)\vert0\rangle+c_{1}(t)\vert1\rangle]\otimes\prod_{n=1}^{N}\vert\Bar{0}\rangle_{n,r}\nonumber\\&+&\vert0\rangle\otimes\sum_{n=1}^{N}\sum_{k}c_{n,k}(t)\vert1_{k}\rangle_{n,r}.
\end{eqnarray}
In the above equation $c_1(t)$ and $c_{n,k}$ are time dependent coefficients of the spin system and the reservoir, respectively. Using Schrodinger's equation Eq.(\ref{1Schrodinger equation}), we obtain the time dependent coefficients in the interaction picture which are governed by the differential equations \cite{PhysRevA.90.062104}:
\begin{eqnarray}
\frac{d}{dt}c_{1}(t)=-i\sum_{n=1}^{N}\sum_{k} h_{n,k}e^{i(w_{eff}-w_{n,k})t}c_{n,k}(t),
\end{eqnarray}
\begin{eqnarray}
\frac{d}{dt}c_{n,k}(t)=-i h_{n,k}^{*}e^{-i(w_{eff}-w_{n,k})t}c_{1}(t).
\end{eqnarray}
We observe that the summation $\sum_{k} |h_{n,k}|^{2}e^{i(w_{eff}-w_{n,k})t}$ appearing in the above equation is the correlation function $s_{n}(t)$ of the $n^{\rm th}$ reservoir. In the limit of a large number of modes, the summation can be obtained in the form of integration in term of spectral density $j_{n}(\omega)$ as
\begin{eqnarray}
s_{n}(t-t_{1})=\int d\omega j_{n}(\omega) \exp{[i (\omega_{eff}-\omega)(t-t_{1})]}.
\label{corelation}
\end{eqnarray}
The coefficient $c_{1}(t)$ can now be expressed as
\begin{eqnarray}
\frac{d}{dt}c_{1}(t)=-\int_{0}^{t} d{t_{1}} c_{1}(t_{1})S(t-t_{1}),
\end{eqnarray}
with $S(t-t_{1})=\sum_{n=1}^{N}s_{n}(t-t_{1})$ and
the spectral density $j_n(\omega)$ is assumed to be Lorentzian form $j_{n}(\omega)=g_{n}\tau_{n}^{2}/\big(2\pi[(\omega_{eff}-\omega-\delta_{n})^2+\tau_{n}^{2}]\big)$,\cite{PhysRevA.96.012105,breuer2002theory}  where $g_n$ is the system-reservoir coupling strength and $\tau_n^{-1}$ is the correlation time of the reservoir. The central frequency of the $n^{\rm th}$ reservoir $\omega_n^c$ is detuned by $\delta_n$ from the $\omega_{eff}$. Considering a simpler case of identical reservoirs and defining
 $g_{n}/ \tau_{n}=g/ \tau$, $\omega_n^c=\omega^c$ and $\delta_n=\delta$, we can obtain the function $c_{1}(t)$ in a compact form as
\begin{eqnarray}\label{non-Markovianity1}
c_{1}(t)=c_{1}(0)e^{-\frac{(\tau-i \delta) t}{2}}\Big[\cosh\biggl(\frac{\kappa_{x} t}{2}\biggr)+\frac{(\tau-i \delta)}{\kappa_{x}} \sinh\biggl(\frac{\kappa_{x} t}{2}\biggr)\Big],\nonumber\\
\end{eqnarray}
with $\kappa_{x}=\sqrt{(\tau-i \delta)^2-2 N g \tau}$ and $\delta=\omega_{eff}-\omega^c$. We observe that the amplitude of $c_1(t)$ is decaying with a rate $\tau^{-1}$ and oscillating with frequency $\delta$.
We can write the dynamics of the NV spin in terms of a reduced density matrix in the basis of $\vert0\rangle$ and  $\vert1\rangle$ by tracing out the reservoirs as shown in Appendix \ref{appendix7},\cite{breuer2002theory}
\begin{equation}
\label{decoherencex}
\rho_{s}(t)=
\begin{pmatrix}
1-\vert c_{1}(t)\vert^2 & c_{0}(0)c_{1}^{*}(t)  \\
 c_{0}^{*}(0)c_{1}(t) & \vert c_{1}(t)\vert^2
\end{pmatrix},\\
\end{equation}
where
\begin{eqnarray}
\vert c_{1}(t)\vert^2 &=&\vert c_{1}(0)\vert^2 e^{-\tau t}\Big[\Big(\cosh\Big(\frac{\kappa' t}{2}\Big)\cos\Big(\frac{\kappa'' t}{2}\Big)\nonumber\\
&+&\frac{\kappa_{1}}{\vert \kappa_{x} \vert^2}\sinh\Big(\frac{\kappa' t}{2}\Big)\cos\Big(\frac{\kappa'' t}{2}\Big)\nonumber \\
&-&\frac{\kappa_{2}}{\vert \kappa_{x} \vert^2}\cosh\Big(\frac{\kappa' t}{2}\Big)\sin\Big(\frac{\kappa'' t}{2}\Big)\Big)^2\nonumber\\
&+&\Big(\sinh\Big(\frac{\kappa' t}{2}\Big)\sin\Big(\frac{\kappa'' t}{2}\Big)\nonumber \\
&+&\frac{\kappa_{1}}{\vert \kappa_{x} \vert^2}\cosh\Big(\frac{\kappa' t}{2}\Big)\sin\Big(\frac{\kappa'' t}{2}\Big)\nonumber\\
&+&\frac{\kappa_{2}}{\vert \kappa_{x} \vert^2}\sinh\Big(\frac{\kappa' t}{2}\Big)\cos\Big(\frac{\kappa'' t}{2}\Big)\Big)^2\Big]
\end{eqnarray}
and $\kappa'$ ($\kappa''$) are real (imaginary) part of $\kappa_{x}$ and $\kappa_{1}= (\tau \kappa'+\delta \kappa'' )$, $\kappa_{2}= ( \delta\kappa'- \tau\kappa'' )$.

In order to ensure that the set of parameters lead us to the dynamics of the system in non-Markovian regime
we calculate the trace distance $\mathcal{D}(\rho_1(t),\rho_2(t))$ between the time evolved states of any two quantum states $\rho_1(t)$ and $\rho_2(t)$ of the system and find the rate of change of the trace distance $F(N,t)$. For all the Markovian processes $F(N,t)$ is less than or equal to zero which means that the information will always flow from the system to the environment.  However, if there exist a pair of initial states for which $F(N,t)$ is positive even for a certain time, the process is said to be non-Markovian. All those times when $F(N,t)$ is positive, the distinguishability of the time evolved pair of initial states increases resulting a back flow of information from the environment to the system. It has been shown \cite{PhysRevA.81.062115} that if  we start with a pair of initial states $|\psi_{1,2}(0)\rangle=\frac{1}{\sqrt{2}}(\vert0\rangle\pm\vert1\rangle)$, the trace distance comes out to be $|c_1(t)|$ and the rate of change of the trace distance turns out to be
\begin{eqnarray}\label{the fundamental property of non-Markovianity1}
F(N,t)=\frac{d|c_1(t)|}{dt}.
\end{eqnarray}
\begin{figure}[t]
  \includegraphics[width=\linewidth,height=2.2in]{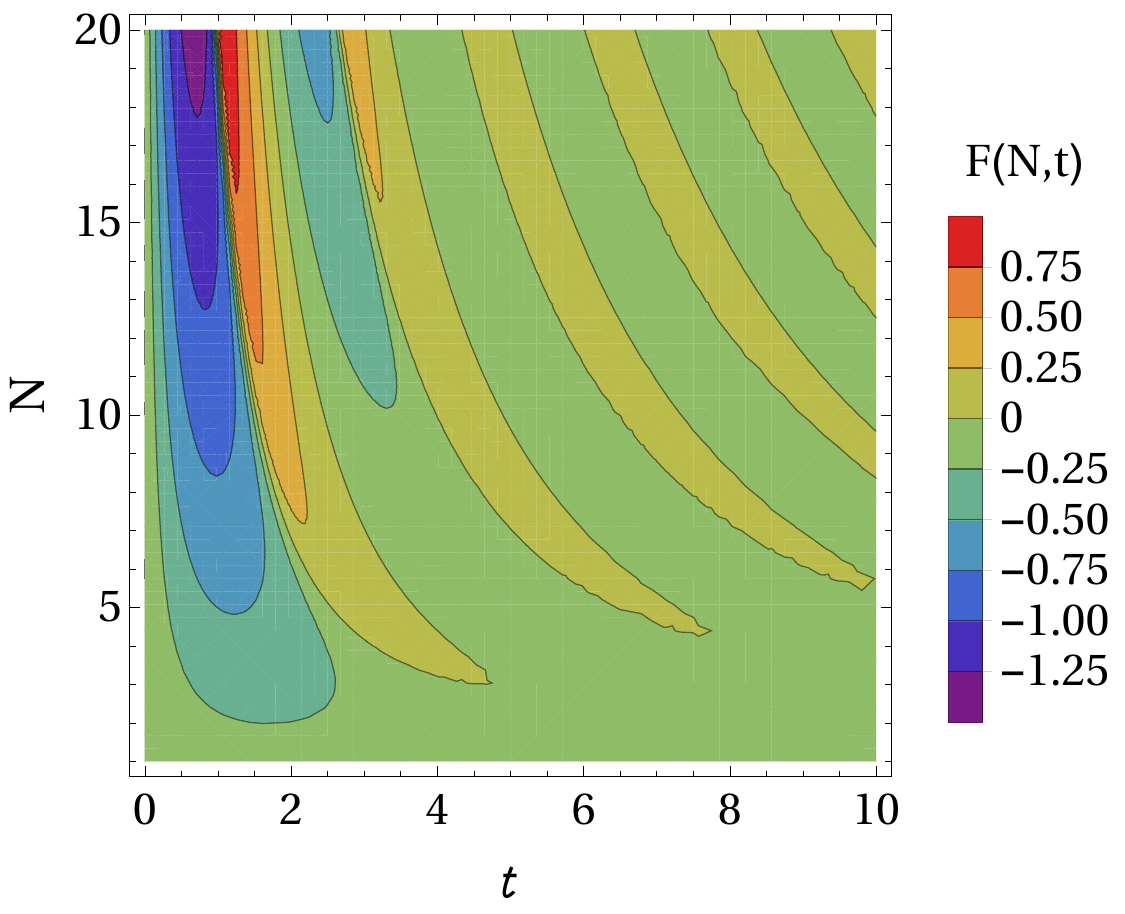}
\caption{ $F(N,t)$ is plotted for different values of number of reservoirs $N$  and time $t$. Weak system-reservoir coupling is considered with parameters $\tau=1$, $g=\frac{3\tau}{8}$ and $\delta=0.2$.}
\label{countour}
 \end{figure}
The full expression of $F(N,t)$ is shown in Appendix \ref{appendix6}. If $F(N,t)$ is larger than zero in a certain time interval, the reservoir displays a non-Markovian behavior \cite{PhysRevLett.103.210401}.

This fact gives a criteria for the choice of parameters $\tau$, $g$, $\delta$ and $N$ such that the reservoirs act as a non-Markovian channel. We can define the quantity $[\tau/2 g+1]=N_{\rm{c}}$ when $\delta=0$. For instance, $ g=\frac{3 \tau}{8} $ gives $N_c=2$. It is worth to note that away from the resonance i.e $\delta \neq 0$, the above definition of $N_c$ will not hold. For example, for $\delta=0.2g$,  the critical number of reservoirs $N_{\rm{c}}=3$ and for $\delta=0.5g$, the critical number of reservoirs becomes $N_{\rm{c}}=4$ for the same choice of $\tau/2g$. In Fig. \ref{countour} we have shown a contour plot of $F(N,t)$ with $N$ and $t$ for parameters $\delta=0.2g$ and $ g=\frac{3 \tau}{8}$. The $F(N,t)>0$ regions form a comb like contours. We see that the amplitude of oscillations of $F(N,t)$ increases with increasing $N$ and regions of prominent $F(N,t)<0$ and $F(N,t)>0$ are visible for high $N$.
 \begin{figure}[t]
  \includegraphics[width=\linewidth,height=2.2in]{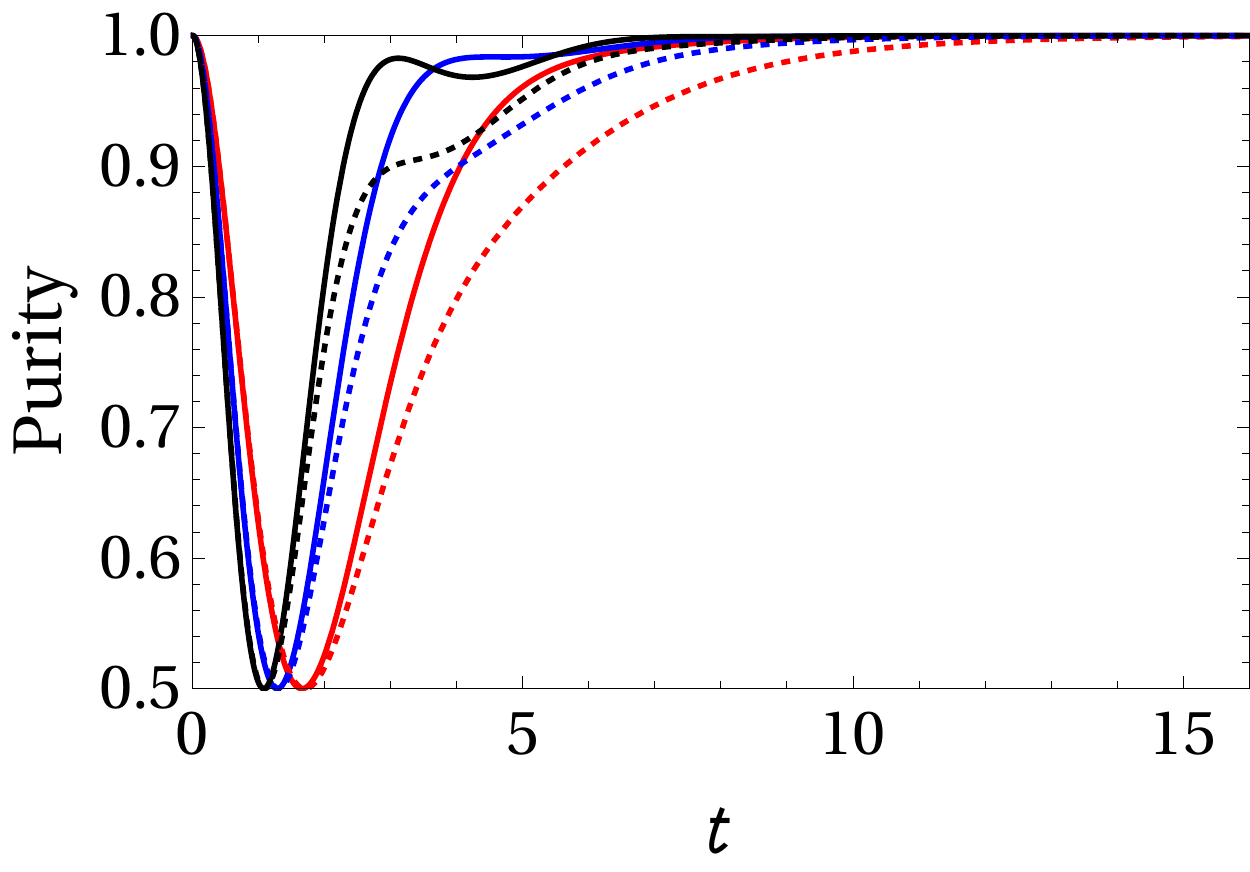}
\caption{Purity of the system coupled with $N$ reservoirs within a weak coupling regime with parameters $\tau=1$ and $g=\frac{3\tau}{8}$, $\delta=0.2$ (\rm{solid}), $\delta=0.5$ (\rm{dotted}) for $c_{0}(0)=0$ and $c_{1}(0)=1$. The red, blue and black lines represent   $N=2,3,4$ cases, respectively.}
\label{puritynonmarkovian}
 \end{figure}

We can calculate the purity of the NV spin using $\mathcal{P}=Tr(\rho_{s}^2(t))$ and get
\begin{eqnarray}
\mathcal{P}&=&1+2 \vert c_{1}(t)\vert^2\Big(\vert c_{1}(t)\vert^2+ \vert c_{0}(0)\vert^2 -1\Big).
\end{eqnarray}

 We can see from Fig. \ref{puritynonmarkovian} that the system loses the purity as time progresses while evolving through the intermediate mixed state, and finally reaching to a pure state. We see that increasing the number of reservoirs leads to a  quicker revival of the pure state and when  system-reservoir coupling is not present i.e., $\tau=0$, the system remains in the pure state. The oscillations in the purity is due to the $\delta$ term in the expression of $c_1(t)$. For the weak system-reservoir coupling i.e., $ g=\frac{3 \tau}{8} $ and $\delta=0.2g$ the system shows Markovian dynamics for $N=2$. In this case if we add more reservoirs the system will show non-Markovian behaviour. In our case we have used the parameters $\tau=1$,  $g = 3\tau/8 $  and $\delta=0.2~(\rm{solid})$ corresponds to $\delta\textless g/\tau$ , $\delta=0.5~(\rm{dotted})$ corresponds to $\delta \textgreater g/\tau$. Increasing the detuning parameter $\delta$ ($\geq g/\tau$) increases the number of reservoirs and enhances the back flow of the information from the environment to the system.  \cite{PhysRevLett.103.210401,PhysRevA.81.062124,PhysRevA.84.032118}
 \section{multilevel dynamics}
\label{multilevel}
The presence of the spin-oscillator coupling term leads to  mixing of the nonlinear oscillator states. We assume that initially the
system is in the $G_{0}$ region. Energy spectrum of the system is such  \cite{Ugulava2005,Chotorlishvili2010,Chotorlishvili2018}, that for a given value of the barrier height $l$,
only several energy levels $E_{n}(l)$ and states belong to the region $G_{0}$. We assume that these two states are neighboring states $ce_{n}$, $se_{n+1}$. Then the computational basis vectors are: $ \vert ce_{n}(\varphi,l) \rangle \otimes  \vert 0 \rangle  $, $ \vert ce_{n}(\varphi,l) \rangle \otimes \vert 1  \rangle $,  $ \vert se_{n+1}(\varphi,l) \rangle \otimes  \vert 0 \rangle  $, and $ \vert se_{n+1}(\varphi,l) \rangle \otimes \vert 1  \rangle $. The Hamiltonian of the system is
\begin{equation}
 \hat{H}=
\begin{pmatrix}
a_{2}+b_{2} & c_{2} & d_{2} & e_{2} \\
c_{2} & a_{2}-b_{2}  & e_{2} & -d_{2}\\
d_{2} &  e_{2} & f_{2}+g_{2} & h_{2} \\
e_{2} & -d_{2} & h_{2} & f_{2}-g_{2}
 \end{pmatrix},
 \label{ham_multilevel}
\end{equation}
where $ a_{2}= a_{n}(l)$ , $ b_{2}=\frac{\omega_{0}}{2} + \frac{1}{2} Q r_{2} \cos {\alpha} $ ,  $ c_{2}=\frac{1}{2} Q  r_{2} \sin {\alpha} $, $ d_{2} =\frac{1}{2}  Q s_{2} \cos {\alpha} $ , $ e_{2} =\frac{1}{2} Q s_{2} \sin {\alpha} $, $f_{2}=b_{n+1}(l)$,$g_{2}=\frac{\omega_{0}}{2} + \frac{1}{2} Q t_{2} \cos {\alpha} $, $h_{2}=\frac{1}{2}Q t_{2} \sin{\alpha} $,
\begin{eqnarray}
r_{2} =[\{\frac{\pi}{2}\left(A_{1}^{(2n+1)}(l)\right)^{2}+
 \pi\sum\limits_{r=0}^{\infty}A_{2r+1}^{(2n+1)}(l)A_{2r+3}^{(2n+1)}(l)\}],\nonumber\\
\end{eqnarray}
\begin{eqnarray}
 t_{2}&=&[\{\frac{\pi}{2}\left(-B_{1}^{(2n+1)}(l)\right)^{2}+
 \pi\sum\limits_{r=0}^{\infty}B_{2r+1}^{(2n+1)}(l)B_{2r+3}^{(2n+1)}(l)\}],\nonumber\\
\end{eqnarray}
$s_{2}=\langle ce_{n}(\varphi,l) \vert \cos(2 \varphi)\vert se_{n+1}(\varphi,l) \rangle.$
\\
The multilevel dynamics of the system in the subgroup $G_{0}$ is far more complicated.
Let us define the initial state of the system in $G_{0}$ region as
 \begin{eqnarray}\label{y}
 \vert \psi(0) \rangle = \vert ce_{n}(\varphi,l) \rangle \otimes \vert 0 \rangle.
 \end{eqnarray}
We use the following ansatz for the wave function in the region $G_{0}$:
\begin{eqnarray}\label{3Schrodinger equation}
&&\vert \psi(t) \rangle=S_{1}(t)\vert ce_{n} (\varphi,l)  \rangle \vert 0 \rangle +S_{2}(t)\vert ce_{n} (\varphi,l)  \rangle \vert 1 \rangle +\nonumber \\
&&S_{3}(t)\vert se_{n+1} (\varphi,l)  \rangle \vert 0 \rangle +S_{4}(t)\vert se_{n+1} (\varphi,l)  \rangle \vert 1 \rangle.
\end{eqnarray}
and solve the Schr\"odinger equation (Eq. (\ref{1Schrodinger equation})) for the Hamiltonian Eq. (\ref{ham_multilevel}) to get coefficients $S_1(t)$, $S_2(t)$, $S_3(t)$ and $S_4(t)$. Using this solution $\psi(t)$
we calculate the density matrix $\rho_{AB}(t)=\vert \psi(t) \rangle \langle \psi(t) \vert$ which is given as
\begin{eqnarray}
\rho_{AB}(t) =
\begin{pmatrix}
\rho_{11} & \rho_{12} & \rho_{13} & \rho_{14} \\
\rho_{21} & \rho_{22}  & \rho_{23} & \rho_{24}\\
\rho_{31} &  \rho_{32} & \rho_{33} & \rho_{34} \\
\rho_{41} & \rho_{42} & \rho_{43} & \rho_{44}
 \end{pmatrix}.
\end{eqnarray}
We  trace out partially the mathematical pendulum part and calculate the reduced density matrix $\hat{\rho}_B(t)$ of the spin part of the system in the region $G_{0}$ as
\begin{eqnarray}\label{reducedlast}
\hat{\rho}_{B}(t)&=&\big(|S_{1}|^{2}+|S_{3}|^{2}\big)|0\rangle\langle0|
+ \big(|S_{2}|^{2}+|S_{4}|^{2}\big)|1\rangle\langle1|\nonumber\\
&+& \big(S_{1}S_{2}^{\ast}+S_{3}S_{4}^{\ast}\big)|0\rangle\langle1|+h.c.,
\end{eqnarray}
The expectation values of longitudinal spin component $\langle \sigma_z\rangle$ and transverse spin components $\langle \sigma_x\rangle$, $\langle \sigma_y\rangle$ are given as
\begin{eqnarray}
 \langle \sigma_{z} \rangle =
 |S_1(t)|^2+|S_3(t)|^2-|S_2(t)|^2-|S_4(t)|^2,
\end{eqnarray}
\begin{eqnarray}
\langle \sigma_{y} \rangle &=& -2Im(S_1(t)S_2^{*}(t)+S_3(t)S_4^{*}(t)),
\end{eqnarray}
and
\begin{eqnarray}
\langle\sigma_{x} \rangle &=&2Re(S_1(t)S_2^{*}(t)+S_3(t)S_4^{*}(t)).
\end{eqnarray}

\begin{figure*}
 \includegraphics[width=\linewidth,height=2.2in]{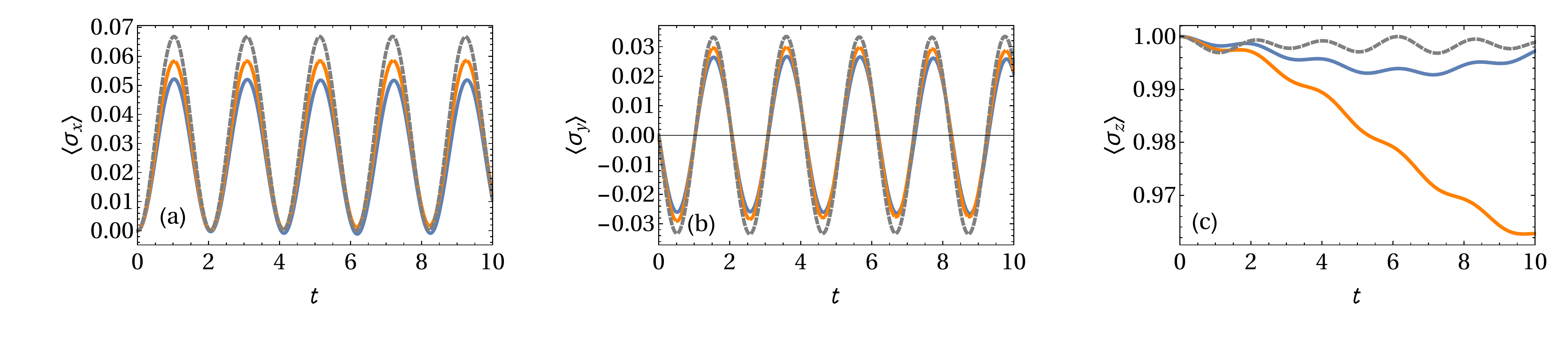}

\caption{(a) Transverse spin component $ \langle \sigma_{x} \rangle $, (b) transverse spin component $ \langle \sigma_{y} \rangle $, and (c) longitudinal spin component $ \langle \sigma_{z} \rangle $ plotted as function of time for the bipartite system $\hat{\rho}_{AB}$ in the region $ G_{0} $ for different multilevel quantum states $n=2,3,4$. The blue (solid), orange (solid) and violet (dashed) lines represent $n=2,l=3.855$,  $n=3,l=7.535$ and  $n=4,l=10.785$ cases, respectively. The values of barrier heights $l$ are chosen to be in the region $G_{0}$ in the vicinity to the transition into the region $G_{-}$. The interaction strength between nonlinear oscillator and NV spin is $ Q = 0.5 $. Time is in the units of $\omega_0^{-1}$.}
\label{figure20}
\end{figure*}

The state given by  Eq. (\ref{reducedlast}) is a multilevel product state as the eigenvalues of the
reduced density matrix are $\rho^{B}_{11}=|\alpha|^{2}+|\beta|^{2}$, $\rho^{B}_{22}=0$
and von Neumann entropy ${\rm Tr}\rho_B\log_2\rho_B$ is zero. The results obtained for the spin dynamics for the multilevel case are shown in the Fig. \ref{figure20} (a), (b) and (c). The transverse spin components $\langle \sigma_x\rangle$, $\langle \sigma_y\rangle$ indicate  switching  and the amplitude of the oscillation increases with increase in the quantum number $n$. In Fig. \ref{figure20} (a) and (b) the behaviour of  $\langle \sigma_x\rangle$, $\langle \sigma_y\rangle$ is the same as that of the Fig. \ref{figure1} (a) and (b) due to a large energy gap between the mathematical pendulum states considered for given $n$ and $l$.  The longitudinal spin part $\langle \sigma_y\rangle$ decays and  more prominent so for certain value of $n$. For instance, in Fig. \ref{figure20} (c), $n=3$ and $l=7.535$ decay due to transition between neighboring states is observed. For this case, we know that initially $|S_1(t)|^2$ is one and all other probabilities are zero. We can show numerically that as time progresses $|S_1(t)|^2$ decreases and  $|S_4(t)|^2$ increases, while $|S_2(t)|^2$ and $|S_3(t)|^2$ are showing a small variation. This indicates a major transition from the initial state $\vert ce_{n} (\varphi,l)  \rangle \vert 0 \rangle$  to the neighbouring state $\vert se_{n+1} (\varphi,l)  \rangle \vert 1 \rangle$.

We note that by steering parameters of the driving term in Eq.(\ref{hamiltonian2}), we can easily achieve the desired value of $l$ from the region $G_{0}$ and switch the NV spin from $|0\rangle$ state to  $|1\rangle$.

\section{Unitary generation of coherence}
\label{unitary_coherence}
The advantage of  NV centers over  other qubit systems is their relatively low decoherence rate.  On the long run, even a slow decoherence leads to substantial effects on the dynamics of the system. Due to  decoherence the off-diagonal elements of the density matrix diminish. The question now is whether this system can evolve to some other state with finite  off-diagonal elements signaling  coherence, a mechanism that may be useful for some coherence-based operations  \cite{gour1,gour2,RevModPhys.89.041003,korzekwa2018coherifying}. 
The mixedness of the state $\rho$ is characterized by the linear entropy given as $[1-Tr(\rho^2)]$. Under unitary evolution the mixedness never changes. It will be important to see whether the coherence can be generated under the unitary evolution.
%
%

Suppose that the system is prepared in a mixed state $\hat{\rho}(0)$ initially corresponding to the Hamiltonian $\hat{H}_{0}$. We consider the unitary evolution of the state $\hat{\rho}(0)$ governed by the operator $\hat{U}=\big(\exp\big(-\frac{i}{\hbar}\int\hat{H}dt\big)\big)$, where
$\hat{H}=\hat{H}_{0}+\hat{V}$. For obtaining closed analytical result, in what follows, we consider a sudden quench of the Zeeman splitting $\frac{1}{2}\Delta \omega_{0}$, i.e., $(\omega_{0},t<0)\rightarrow (\omega_{0}+\Delta \omega_{0},t>0)$.
Let  us begin by assuming the system to be  initially in the region $G_{0}$  and  prepared in the mixed state as:

\begin{eqnarray}\label{mixed_state1}
\hat{\rho}(0)=p_{1}|\phi_{1}\rangle\langle\phi_{1}|+p_{2}|\phi_{2}\rangle\langle\phi_{2}|.
\end{eqnarray}
Where $|\phi_{1}\rangle$ and $|\phi_{2}\rangle$ are the eigenstates of Hamiltonian in $G_{0}$ region. 
Following the idea put forward  in Refs.\cite{akallush2019, Koslof}, we present the total Hamiltonian  Eq. (\ref{total1}) after the quench in the form:
\begin{eqnarray}\label{timedependentHamiltonian}
&&\hat{H}=\hat{H}_{0}+\hat{V},
\end{eqnarray}
where,
\begin{eqnarray}
&& \hat{H}_{0}=\hat{H}_{m}+\hat{H}_{s}+Q\cos(\varphi)\hat{S}_{z}\nonumber\\
&&\hat{V}=\frac{1}{2}\Delta \omega_{0}\sigma_{z}.
\end{eqnarray}
We note that $\hat{S}_{z}$ contains the raising and lowering operators $\hat{\sigma}_{+},~\hat{\sigma}_{-}$ and therefore the commutator is not zero $[\hat{H}_{s},\hat{V}]\neq 0$.
We exploit the relative entropy as an entropic measure for coherence:
\begin{eqnarray}\label{divergence}
&&\mathcal{C}\big(\hat{\rho}(t)|\hat{\rho}_{d}\big)=Tr\{\hat{\rho}(t)\ln\hat{\rho}(t)-\hat{\rho}(t)\ln\hat{\rho}_{d}\}.
\end{eqnarray}
Here $\hat{\rho}_{d}$ is the diagonal part of the propagated density matrix $\hat{\rho}(t)=\exp(\frac{i}{\hbar} \hat{H}t) \hat{\rho}(0)\exp(-\frac{i}{\hbar} \hat{H}t)$.
The larger is the departure from the $\hat{\rho}_d$, larger is the relative entropy $\mathcal{C}\big(\hat{\rho}(t)|\hat{\rho}_{d}\big)$. This departure
from the $\hat{\rho}_d$ is quantified by the non-zero off-diagonal part of $\hat{\rho}(t)$. Since we start from an incoherent state $\hat{\rho}(0)$, the non zero off-diagonal elements of $\hat{\rho}(t)$ signal the generation of coherence.  

As for the  relation between coherence and purity we refer to Ref. [\onlinecite{rastegin2016quantum}] and Ref. [\onlinecite{singh2015maximally}] stating 
\begin{eqnarray}\label{coherence}
&&\mathcal{C}\big(\hat{\rho}(t)|\hat{\rho}_{d}\big)\leq\sqrt{2\mathcal{P}-1}.
\end{eqnarray}
Eq.(\ref{coherence}) shows that there is an upper bound of the coherence quantified through the purity
$\mathcal{P}$. This means that during the unitary evolution, coherence can be changed  even though purity is invariant under unitary evolution $\mathcal{P}(\hat{\rho}(t))=\mathcal{P}(\hat{\rho}(0))$.
For  incoherent unitaries coherence is constant \cite{PhysRevLett.113.140401}.
The time evolved state $\rho(t)$ for the initial state given by  Eq. (\ref{mixed_state1}) and the Hamiltonian given by Eq. (\ref{timedependentHamiltonian}) is calculated as 
\begin{widetext}
\begin{eqnarray}\label{timedensitymatrixes}
\hat{\rho}(t)&=& \left(p_{1}|\langle\psi_{1}|\phi_{1}\rangle|^{2}+p_{2}|\langle\psi_{1}|\phi_{2}\rangle|^{2}\right)|\psi_{1}\rangle\langle\psi_{1}|
+\left( p_{1}|\langle\psi_{2}|\phi_{1}\rangle|^{2}+p_{2}|\langle\psi_{2}|\phi_{2}\rangle|^{2}\right)|\psi_{2}\rangle\langle\psi_{2}| \nonumber\\
&+&\exp\left(-\frac{i}{\hbar}\left(E_{1}-E_{2}\right)t\right)\left(p_{1}\langle\phi_{1}|\psi_{2}\rangle\langle\psi_{1}|\phi_{1}\rangle +
p_{2}\langle\phi_{2}|\psi_{2}\rangle\langle\psi_{1}|\phi_{2}\rangle\right)|\psi_{1}\rangle\langle\psi_{2}|\nonumber\\
&+&\exp\left(\frac{i}{\hbar}\left(E_{1}-E_{2}\right)t\right)\left(p_{1}\langle\phi_{1}|\psi_{1}\rangle\langle\psi_{2}|\phi_{1}\rangle+
p_{2}\langle\phi_{2}|\psi_{1}\rangle\langle\psi_{2}|\phi_{2}\rangle\right)|\psi_{2}\rangle\langle\psi_{1}|.
\end{eqnarray}
\end{widetext}
Here $|\phi_{1,2}\rangle$ are the eigenvectors and $E_{1,n}(l),E_{2,n}(l)$ are the eigenvalues of $\hat{H}_{0}(l)$ given in the explicit form as
\begin{eqnarray}\label{are the eigenfunctions of1}
&&|\phi_{1}\rangle=|ce_{n}(l)\rangle\otimes\left(\alpha_{1}|1\rangle+\beta_{1}|0\rangle\right),\nonumber\\
&&|\phi_{2}\rangle=|ce_{n}(l)\rangle\otimes\left(\beta_{1}|1\rangle-\alpha_{1}|0\rangle\right), \nonumber\\
&&\alpha_{1}=1/\sqrt{\lambda^{2}+1},~\beta_{1}=\lambda/\sqrt{\lambda^{2}+1},\nonumber\\
&&\lambda=(b+\sqrt{b^{2}+c^{2}})/c,\nonumber\\
&& E_{1,n}(l)=a + \sqrt{b^{2} + c^{2}}, E_{2,n}(l)=a - \sqrt{b^{2} + c^{2}},
\end{eqnarray}
and coefficients $a$, $b$ and $c$ are already defined in section IV.
We  write the Hamiltonian after the quench $\hat{H}$ in the diagonal basis of $H_0$ as
\begin{equation}
\label{Hamiltonian_after_quench}
\hat{H}=
\begin{pmatrix}
E_{1,n}(l)+J &Y  \\
Y & E_{2,n}(l)-J)
\end{pmatrix},\\
\end{equation}
where $J$ and $Y$ are given as: 
\begin{eqnarray}
&&J=-\Delta \omega_{0}(\alpha_{1}^{2}-\beta_{1}^{2}),\nonumber\\
&&Y=-2\Delta\omega_{0}\alpha_{1}\beta_{1}.
\end{eqnarray}
The eigenvectors of $\hat{H}$ take explicit form as:
\begin{eqnarray}\label{are the eigenfunctions of2}
&&|\psi_{1}\rangle=\zeta_{1}|\phi_{1}\rangle+\zeta_{2}|\phi_{2}\rangle,\nonumber\\
&&|\psi_{2}\rangle=\zeta_{2}|\phi_{1}\rangle-\zeta_{1}|\phi_{2}\rangle.
\end{eqnarray}
The coefficients $\zeta_1$, $\zeta_2$, eigenvalues of $\hat{H}$ and the rest of the information are presented in the Appendix \ref{appendix_F}.
Taking into account Eq.(\ref{timedensitymatrixes})-Eq.(\ref{are the eigenfunctions of2}) we rewrite the propagated density matrix in the more compact form

\begin{eqnarray}\label{compact}
&&\hat{\rho}(t)=(p_{1}\zeta_{1}^{2}+p_{2}\zeta_{2}^{2})|\psi_{1}\rangle\langle\psi_{1}|+
(p_{1}\zeta_{2}^{2}+p_{2}\zeta_{1}^{2})|\psi_{2}\rangle\langle\psi_{2}|\nonumber\\
&&+\exp(-i\omega_{12}t)\zeta_{1}\zeta_{2}(p_{1}-p_{2})|\psi_{1}\rangle\langle\psi_{2}|\nonumber\\
&&+\exp(i\omega_{12}t)\zeta_{1}\zeta_{2}(p_{1}-p_{2})|\psi_{2}\rangle\langle\psi_{1}|.
\end{eqnarray}

\begin{figure}
 \includegraphics[width=\linewidth,height=2.2in]{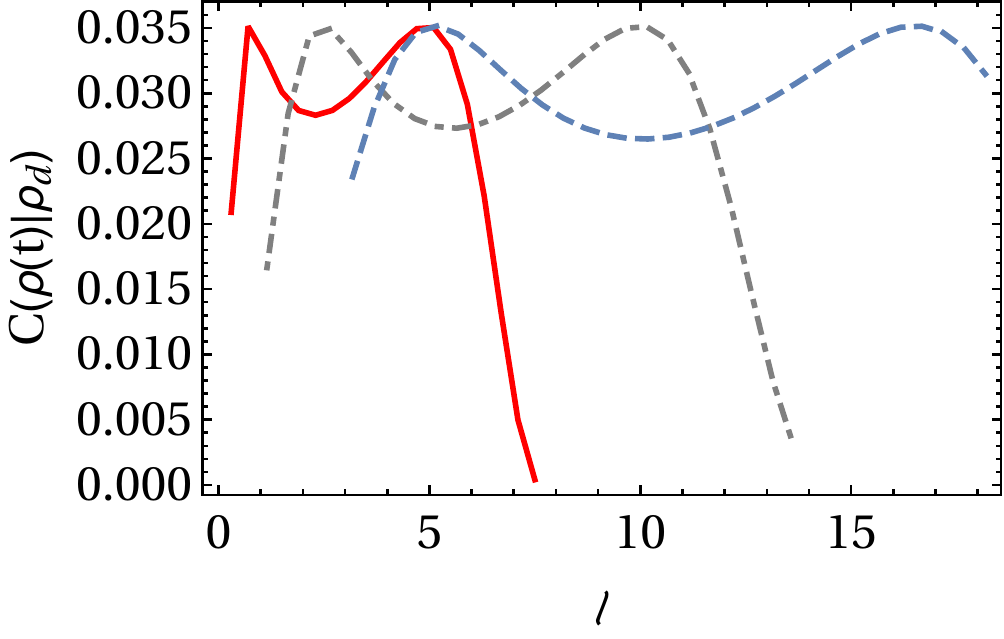}
\caption{The Coherence $ \mathcal{C}(\rho(t)|\rho_{d}) $  plotted for the bipartite system  in the region $ G_{0} $ for different quantum states $n=2,3,4$ and corresponding barrier height $l$ in $G_{0}$ region . The red (solid), violet (Dotted-Dashed) and blue (Dashed) lines represent $n=2$,  $n=3$ and  $n=4$ cases, respectively. The parameter used for the plot is $p_{1}=0.9$ and $p_{2}=0.1$, $\Delta \omega_{0}=0.8$ and $\omega_{0}=1$, $ Q = 5 $.}
\label{divergence_fig}
 \end{figure}

Note that the evolved density matrix $\hat{\rho}(t)$, Eq.(\ref{compact}) is not diagonal in the basis Eq.(\ref{are the eigenfunctions of2}) of the quenched Hamiltonian Eq. (\ref{Hamiltonian_after_quench}). Diagonalizing  the evolved density matrix $\hat{\rho}(t)$ (Eq.(\ref{compact})) we obtain the following eigenvectors
\begin{eqnarray}\label{are the eigenfunctions of3}
&&|\rho_{1}\rangle=\nu_{1}|\psi_{1}\rangle+\nu_{2}|\psi_{2}\rangle,\nonumber\\
&&|\rho_{2}\rangle=\nu_{2}|\psi_{1}\rangle-\nu_{1}|\psi_{2}\rangle.
\end{eqnarray}
The coefficients $\nu_1$,  $\nu_2$ and eigenvalues are presented in the appendix \ref{appendix_F}.
Taking into account Eq.(\ref{timedensitymatrixes})-Eq.(\ref{are the eigenfunctions of3}), for the
quantum coherence Eq.(\ref{divergence}) we derive

\begin{eqnarray}\label{divergence2}
&&\mathcal{C}\big(\hat{\rho}(t)|\hat{\rho}_{d}\big)=p_{1}\ln(p_{1})+p_{2}\ln(p_{2})-
p_{1}|\langle\rho_{1}|\psi_{1}\rangle|^{2}\ln(p_{1}\zeta_{1}^{2}\nonumber\\&&+p_{2}\zeta_{2}^{2})-p_{2}|\langle\rho_{2}|\psi_{2}\rangle|^{2}\ln(p_{1}\zeta_{2}^{2}+p_{2}\zeta_{1}^{2})-
p_{1}|\langle\rho_{1}|\psi_{2}\rangle|^{2}\nonumber\\&&\ln(p_{1}\zeta_{2}^{2}+p_{2}\zeta_{1}^{2})-
p_{2}|\langle\rho_{2}|\psi_{1}\rangle|^{2}\ln(p_{1}\zeta_{1}^{2}+p_{2}\zeta_{2}^{2}),
\end{eqnarray}
or after using trigonometric parametrization (see appendix \ref{appendix_F}) in the explicit form:
\begin{eqnarray}\label{divergence3}
&&\mathcal{C}\big(\hat{\rho}(t)|\hat{\rho}_{d}\big)=p_{1}\ln(p_{1})+p_{2}\ln(p_{2})\nonumber\\&&-
\left(p_{1}\sin^{2}\Theta+p_{2}\cos^{2}\Theta\right)\ln\left(p_{1}\sin^{2}\Theta+p_{2}\cos^{2}\Theta\right)-
\nonumber\\
&&\left(p_{1}\cos^{2}\Theta+p_{2}\sin^{2}\Theta\right)\ln\left(p_{1}\cos^{2}\Theta +p_{2}\sin^{2}\Theta\right).
\end{eqnarray}
All the parameters from Eq.(\ref{divergence3}) in the explicit form are presented in the Appendix \ref{appendix_F}.The hallmark of quantum
chaos is the enhanced fluctuations which is inherent in
this system, for more details see Ref. [\onlinecite{Chotorlishvili2018}]. Therefore, the
coherence $\mathcal{C}\big(\hat{\rho}(t)|\hat{\rho}_{d}\big)$
is rather sensitive with respect to the initial state and values of the parameters.
Further, it is worth  noting that the dynamical chaos emerges in the vicinity of the classical separatrix. In the quantum case, this region corresponds to crossover region between $G_0$ and $G_-$. In particular, this corresponds to 
the selective choice of the barrier height for each quantum state, i.e. $(n=2, l= 0.3-7.51)$, $(n=3, l =1.15-13.93)$, and $(n=4, l=3.18-18.4)$. We plot the coherence from Eq.(\ref{divergence3}) as a function of the barrier height $l$ as shown in Fig. \ref{divergence_fig}. As we see the generation of coherence is maximal when $l$ is chosen from the chaotic region.

\begin{figure}
\includegraphics[width=\linewidth,height=2.2in]{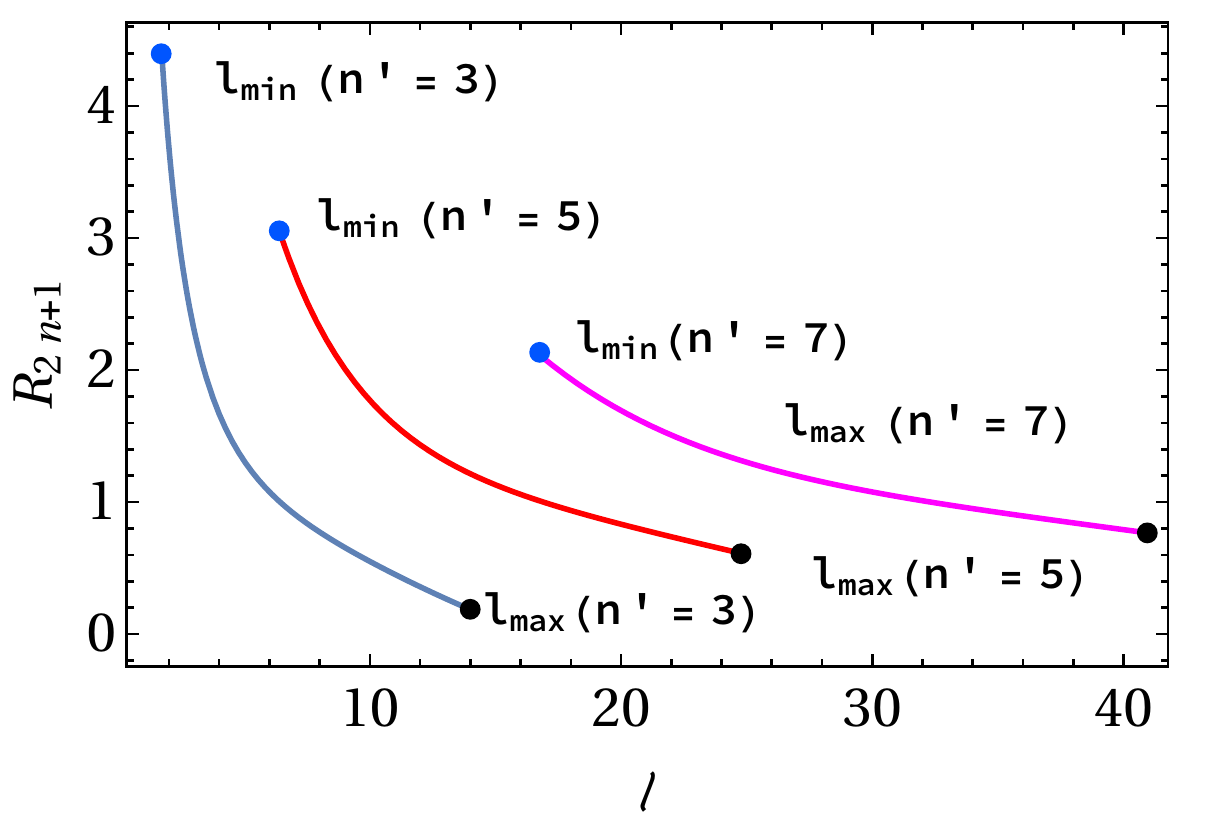}
\caption{The quantum distance $R_{2n+1}$ from the classically chaotic region plotted as a function of the barrier height $l$. The values of barrier heights $l$ are chosen from the region $G_{0}$ in the vicinity to the transition into the region $G_{-}$, i.e. in the vicinity to classical homoclinic tangle.}
\label{figr}
\end{figure}

The result can be explained as follows: the analytical solution of the classical mathematical pendulum Eq.(\ref{total}) has a bifurcation features:
\begin{equation}
\triangle I_{+}=\sqrt{(E+U)/\omega'}dn\big[\omega'\sqrt{(E+U)/\omega'}t,k\big],
\label{solutionclassical1}
\end{equation}
for $E>U$ and
\begin{equation}
\triangle I_{-}=\sqrt{(E+U)/\omega'}cn\big[\omega'\sqrt{(E+U)/\omega'}t,1/k\big],
\label{solutionclassical2}
\end{equation}
for $E<U$. Here $cn(\cdots),~dn(\cdots)$ are Jacobi elliptic functions and parameter is defined as follows $k=\sqrt{2U/(E+U)}$. When $k\rightarrow 1$ in the system occurs bifurcation and solutions take a form of instanton:
\begin{equation}
\Delta I_{+}=I_{-}=\frac{\sqrt{2U/\omega'}}{\cosh(\sqrt{2U\omega'}t)}.
\label{instanton}
\end{equation}
Any small perturbation applied to the system in the vicinity of the bifurcation region
$k\rightarrow 1$ leads to the formation of dynamical chaos and homoclinic tangle.  \cite{zaslavsky2007physics}
The width of the homoclinic tangle read:
\begin{equation}
\frac{|E-U/\omega'|}{U}\preceq \exp\left(-\pi\frac{\nu\sqrt{\omega'}}{\sqrt{U}}\right).
\label{widthtangle}
\end{equation}
Phase trajectories of the system passing through the homoclinic tangle have limited memory, meaning that the information about the initial conditions is gradually lost. By analogy with the classical case, we presume that the quantum systems evolved through the region of quantum chaos have limited memory and weakly depend on the initial state. Therefore, quantum chaos can sustain the generation of coherence from the mixed initial state.
We aim to explore the signature of the quantum chaos and define quantum rescaled distance from the homoclinic tangle
for the odd states $n'=2n+1$ as
\begin{eqnarray}
\mathcal{R}_{2n+1}&=&\frac{1}{l}\left\{a_{2n+1}(l)-\langle ce_{2n+1}(l,\varphi)|V(l,\varphi)|ce_{2n+1}(l,\varphi)\rangle\right\}\nonumber\\&=&
\frac{1}{l}a_{2n+1}(l)-\frac{1}{4}\sum\limits_{r=0}^{\infty}A_{2r+1}^{2m+1}\left(A_{2r+3}^{2m+1}+A_{2r-1}^{2m+1}\right).
\label{Homoclinic tangle}
\end{eqnarray}
Here $V(l,\varphi)=2l\cos2\varphi$ and $a_{2n+1}(l)$ is the Mathieu characteristic.
Direct calculation of the distance from the Homoclinic tangle for the different states
(see Fig. \ref{divergence_fig} and Fig. \ref{figr}) shows that
$\mathcal{R}_{2n+1}$ is larger for the states with minimal coherence, while the initial mixed states favourable to the generation of coherence belong to the chaotic region.
\section{Conclusion}
\label{summary}
The study is focused on a paradigmatic model of NEMS hybrid system: nonlinear oscillator coupled to the spin-1/2 system. Of interest is the spin dynamics in the region where energy spectrum of the system  depends  on  the  height of  the potential  barrier, and  contains  degenerate and non-degenerate  areas corresponding  to  the  different  symmetry  subgroups.
 Varying  the height of the potential barrier switches the symmetry subgroup from degenerate to non-degenerate areas. The isolated system is always in a pure state.  The open quantum system initially prepared in a pure state evolves through the intermediate mixed state and finally reaches to a pure state. The dynamics of the longitudinal spin component allows for a fast switching for strong coupling between NV spin and NEMS. However, the coupling of the system of NEMS and NV spin to the environment leads to slower switching.  We have also investigated the effects of non-Markovian noise originating due to the spin bath $^{13}C$ nuclei.
 Investigating the divergence $C(\rho(t)|\rho_{d})$
which quantifies the generation of coherence, we find that
the generation of coherence through the unitary transformation is  efficient if the system  is prepared initially
in the chaotic region.
\appendix
\section{Matrix elements  in $G_-$, $G_0$ and $G_+$ regions}
\label{appendix1}
The matrix elements corresponding to  Eq. (\ref{allG-ham}) in the region $G_-$:
\begin{eqnarray}
&&A_{11} = E_n(l)+\omega_0/2+\frac{Q}{2}[\{\frac{\pi}{2}\left(A_{1}^{(2n+1)}(l)\right)^{2}\nonumber\\&&+
 \pi\sum\limits_{r=0}^{\infty}A_{2r+1}^{(2n+1)}(l)A_{2r+3}^{(2n+1)}(l)\}+\{\frac{\pi}{2}\left(-B_{1}^{(2n+1)}(l)\right)^{2}\nonumber\\&&+
 \pi\sum\limits_{r=0}^{\infty}B_{2r+1}^{(2n+1)}(l)B_{2r+3}^{(2n+1)}(l)\}  ] \frac{1}{2} \cos\alpha,
\end{eqnarray}
\begin{eqnarray}
&& A_{12} = \frac{Q}{2}[\{\frac{\pi}{2}\left(A_{1}^{(2n+1)}(l)\right)^{2}\nonumber\\&&+
 \pi\sum\limits_{r=0}^{\infty}A_{2r+1}^{(2n+1)}(l)A_{2r+3}^{(2n+1)}(l)\}+\{\frac{\pi}{2}\left(-B_{1}^{(2n+1)}(l)\right)^{2}\nonumber\\&&+
 \pi\sum\limits_{r=0}^{\infty}B_{2r+1}^{(2n+1)}(l)B_{2r+3}^{(2n+1)}(l)\}  ] \frac{1}{2} \sin\alpha,
\end{eqnarray}
\begin{eqnarray}
&&A_{13} = \frac{Q}{2}[\{\frac{\pi}{2}\left(A_{1}^{(2n+1)}(l)\right)^{2}\nonumber\\&&+
 \pi\sum\limits_{r=0}^{\infty}A_{2r+1}^{(2n+1)}(l)A_{2r+3}^{(2n+1)}(l)\}-\{\frac{\pi}{2}\left(-B_{1}^{(2n+1)}(l)\right)^{2}\nonumber\\&&+
 \pi\sum\limits_{r=0}^{\infty}B_{2r+1}^{(2n+1)}(l)B_{2r+3}^{(2n+1)}(l)\}  ] \frac{1}{2} \cos\alpha,
\end{eqnarray}
\begin{eqnarray}
&& A_{14} = \frac{Q}{2}[\{\frac{\pi}{2}\left(A_{1}^{(2n+1)}(l)\right)^{2}\nonumber\\&&+
 \pi\sum\limits_{r=0}^{\infty}A_{2r+1}^{(2n+1)}(l)A_{2r+3}^{(2n+1)}(l)\}-\{\frac{\pi}{2}\left(-B_{1}^{(2n+1)}(l)\right)^{2}\nonumber\\&&+
 \pi\sum\limits_{r=0}^{\infty}B_{2r+1}^{(2n+1)}(l)B_{2r+3}^{(2n+1)}(l)\}  ] \frac{1}{2} \sin\alpha,
\end{eqnarray}
\begin{eqnarray}
&&A_{21} = \frac{Q}{2}[\{\frac{\pi}{2}\left(A_{1}^{(2n+1)}(l)\right)^{2}\nonumber\\&&+
 \pi\sum\limits_{r=0}^{\infty}A_{2r+1}^{(2n+1)}(l)A_{2r+3}^{(2n+1)}(l)\}+\{\frac{\pi}{2}\left(-B_{1}^{(2n+1)}(l)\right)^{2}\nonumber\\&&+
 \pi\sum\limits_{r=0}^{\infty}B_{2r+1}^{(2n+1)}(l)B_{2r+3}^{(2n+1)}(l)\}  ] \frac{1}{2} \sin\alpha,
\end{eqnarray}
\begin{eqnarray}
&&A_{22}=E_n(l)-\omega_0/2+\frac{Q}{2}[\{\frac{\pi}{2}\left(A_{1}^{(2n+1)}(l)\right)^{2}\nonumber\\&&+
 \pi\sum\limits_{r=0}^{\infty}A_{2r+1}^{(2n+1)}(l)A_{2r+3}^{(2n+1)}(l)\}+\{\frac{\pi}{2}\left(-B_{1}^{(2n+1)}(l)\right)^{2}\nonumber\\&&+
 \pi\sum\limits_{r=0}^{\infty}B_{2r+1}^{(2n+1)}(l)B_{2r+3}^{(2n+1)}(l)\}  ]\frac{1}{2} (-\cos\alpha) ,
\end{eqnarray}
\begin{eqnarray}
&&A_{23}=\frac{Q}{2}[\{\frac{\pi}{2}\left(A_{1}^{(2n+1)}(l)\right)^{2}\nonumber\\&&+
 \pi\sum\limits_{r=0}^{\infty}A_{2r+1}^{(2n+1)}(l)A_{2r+3}^{(2n+1)}(l)\}-\{\frac{\pi}{2}\left(-B_{1}^{(2n+1)}(l)\right)^{2}\nonumber\\&&+
 \pi\sum\limits_{r=0}^{\infty}B_{2r+1}^{(2n+1)}(l)B_{2r+3}^{(2n+1)}(l)\}  ]\frac{1}{2} \sin\alpha ,
\end{eqnarray}
\begin{eqnarray}
&&A_{24}=\frac{Q}{2}[\{\frac{\pi}{2}\left(A_{1}^{(2n+1)}(l)\right)^{2}\nonumber\\&&+
 \pi\sum\limits_{r=0}^{\infty}A_{2r+1}^{(2n+1)}(l)A_{2r+3}^{(2n+1)}(l)\}-\{\frac{\pi}{2}\left(-B_{1}^{(2n+1)}(l)\right)^{2}\nonumber\\&&+
 \pi\sum\limits_{r=0}^{\infty}B_{2r+1}^{(2n+1)}(l)B_{2r+3}^{(2n+1)}(l)\}  ]\frac{1}{2} (-\cos\alpha) ,
\end{eqnarray}
\begin{eqnarray}
&&A_{31}=\frac{Q}{2}[\{\frac{\pi}{2}\left(A_{1}^{(2n+1)}(l)\right)^{2}\nonumber\\&&+
 \pi\sum\limits_{r=0}^{\infty}A_{2r+1}^{(2n+1)}(l)A_{2r+3}^{(2n+1)}(l)\}-\{\frac{\pi}{2}\left(-B_{1}^{(2n+1)}(l)\right)^{2}\nonumber\\&&+
 \pi\sum\limits_{r=0}^{\infty}B_{2r+1}^{(2n+1)}(l)B_{2r+3}^{(2n+1)}(l)\}  ]\frac{1}{2} \cos\alpha ,
\end{eqnarray}
\begin{eqnarray}
&&A_{32}=\frac{Q}{2}[\{\frac{\pi}{2}\left(A_{1}^{(2n+1)}(l)\right)^{2}\nonumber\\&&+
 \pi\sum\limits_{r=0}^{\infty}A_{2r+1}^{(2n+1)}(l)A_{2r+3}^{(2n+1)}(l)\}-\{\frac{\pi}{2}\left(-B_{1}^{(2n+1)}(l)\right)^{2}\nonumber\\&&+
 \pi\sum\limits_{r=0}^{\infty}B_{2r+1}^{(2n+1)}(l)B_{2r+3}^{(2n+1)}(l)\}  ]\frac{1}{2} \sin\alpha ,
\end{eqnarray}

\begin{eqnarray}
&&A_{33}=E_n(l)+\omega_0/2+\frac{Q}{2}[\{\frac{\pi}{2}\left(A_{1}^{(2n+1)}(l)\right)^{2}\nonumber\\&&+
 \pi\sum\limits_{r=0}^{\infty}A_{2r+1}^{(2n+1)}(l)A_{2r+3}^{(2n+1)}(l)\}+\{\frac{\pi}{2}\left(-B_{1}^{(2n+1)}(l)\right)^{2}\nonumber\\&&+
 \pi\sum\limits_{r=0}^{\infty}B_{2r+1}^{(2n+1)}(l)B_{2r+3}^{(2n+1)}(l)\}  ] \frac{1}{2} \cos\alpha ,
\end{eqnarray}
\begin{eqnarray}
&&A_{34}=\frac{Q}{2}[\{\frac{\pi}{2}\left(A_{1}^{(2n+1)}(l)\right)^{2}\nonumber\\&&+
 \pi\sum\limits_{r=0}^{\infty}A_{2r+1}^{(2n+1)}(l)A_{2r+3}^{(2n+1)}(l)\}+\{\frac{\pi}{2}\left(-B_{1}^{(2n+1)}(l)\right)^{2}\nonumber\\&&+
 \pi\sum\limits_{r=0}^{\infty}B_{2r+1}^{(2n+1)}(l)B_{2r+3}^{(2n+1)}(l)\}  ]\frac{1}{2} \sin\alpha ,
\end{eqnarray}
\begin{eqnarray}
&&A_{41}=\frac{Q}{2}[\{\frac{\pi}{2}\left(A_{1}^{(2n+1)}(l)\right)^{2}\nonumber\\&&+
 \pi\sum\limits_{r=0}^{\infty}A_{2r+1}^{(2n+1)}(l)A_{2r+3}^{(2n+1)}(l)\}-\{\frac{\pi}{2}\left(-B_{1}^{(2n+1)}(l)\right)^{2}\nonumber\\&&+
 \pi\sum\limits_{r=0}^{\infty}B_{2r+1}^{(2n+1)}(l)B_{2r+3}^{(2n+1)}(l)\}  ]\frac{1}{2} \sin\alpha ,
\end{eqnarray}
\begin{eqnarray}
&&A_{42}=\frac{Q}{2}[\{\frac{\pi}{2}\left(A_{1}^{(2n+1)}(l)\right)^{2}\nonumber\\&&+
 \pi\sum\limits_{r=0}^{\infty}A_{2r+1}^{(2n+1)}(l)A_{2r+3}^{(2n+1)}(l)\}-\{\frac{\pi}{2}\left(-B_{1}^{(2n+1)}(l)\right)^{2}\nonumber\\&&+
 \pi\sum\limits_{r=0}^{\infty}B_{2r+1}^{(2n+1)}(l)B_{2r+3}^{(2n+1)}(l)\}  ]\frac{1}{2} (-\cos\alpha) ,
\end{eqnarray}
\begin{eqnarray}
&&A_{43}=\frac{Q}{2}[\{\frac{\pi}{2}\left(A_{1}^{(2n+1)}(l)\right)^{2}\nonumber\\&&+
 \pi\sum\limits_{r=0}^{\infty}A_{2r+1}^{(2n+1)}(l)A_{2r+3}^{(2n+1)}(l)\}+\{\frac{\pi}{2}\left(-B_{1}^{(2n+1)}(l)\right)^{2}\nonumber\\&&+
 \pi\sum\limits_{r=0}^{\infty}B_{2r+1}^{(2n+1)}(l)B_{2r+3}^{(2n+1)}(l)\}  ]\frac{1}{2} \sin\alpha ,
\end{eqnarray}
\begin{eqnarray}
&&A_{44}= E_n(l)-\omega_0/2+\frac{Q}{2}[\{\frac{\pi}{2}\left(A_{1}^{(2n+1)}(l)\right)^{2}\nonumber\\&&+
 \pi\sum\limits_{r=0}^{\infty}A_{2r+1}^{(2n+1)}(l)A_{2r+3}^{(2n+1)}(l)\}+\{\frac{\pi}{2}\left(-B_{1}^{(2n+1)}(l)\right)^{2}\nonumber\\&&+
 \pi\sum\limits_{r=0}^{\infty}B_{2r+1}^{(2n+1)}(l)B_{2r+3}^{(2n+1)}(l)\}  ] \frac{1}{2} (-\cos\alpha) ,
\end{eqnarray}

The matrix elements corresponding to  Eq. (\ref{allG-ham}) in the region $G_{0}$:
\begin{eqnarray}
&&A_{11}= a_n(l)+\omega_0/2+Q [\{\frac{\pi}{2}\left(A_{1}^{(2n+1)}(l)\right)^{2}\nonumber\\&&+
 \pi\sum\limits_{r=0}^{\infty}A_{2r+1}^{(2n+1)}(l)A_{2r+3}^{(2n+1)}(l)\}]\frac{1}{2} \cos\alpha ,
\end{eqnarray}
\begin{eqnarray}
 &&A_{12}=  Q [\{\frac{\pi}{2}\left(A_{1}^{(2n+1)}(l)\right)^{2}\nonumber\\&&+
 \pi\sum\limits_{r=0}^{\infty}A_{2r+1}^{(2n+1)}(l)A_{2r+3}^{(2n+1)}(l)\}]\frac{1}{2} \sin\alpha ,
\end{eqnarray}
\begin{eqnarray}
 &&A_{21} =Q[\{\frac{\pi}{2}\left(A_{1}^{(2n+1)}(l)\right)^{2}\nonumber\\&&+
 \pi\sum\limits_{r=0}^{\infty}A_{2r+1}^{(2n+1)}(l)A_{2r+3}^{(2n+1)}(l)\}] \frac{1}{2} \sin\alpha ,
\end{eqnarray}
\begin{eqnarray}
&& A_{22}=a_n(l)-\omega_0/2 +Q[\{\frac{\pi}{2}\left(A_{1}^{(2n+1)}(l)\right)^{2}\nonumber\\&&+
 \pi\sum\limits_{r=0}^{\infty}A_{2r+1}^{(2n+1)}(l)A_{2r+3}^{(2n+1)}(l)\}] \frac{1}{2} (-\cos\alpha) ,\nonumber\\
\end{eqnarray}

The matrix elements corresponding to  Eq. (\ref{allG-ham})in the region $G_{+}$:
\begin{eqnarray}
&&A_{11} = E_n(l)+\omega_0/2+\frac{Q}{2}[\{\frac{\pi}{2}\left(A_{1}^{(2n+1)}(l)\right)^{2}\nonumber\\&&+
 \pi\sum\limits_{r=0}^{\infty}A_{2r+1}^{(2n+1)}(l)A_{2r+3}^{(2n+1)}(l)\}\nonumber\\&&+
 \{\pi\sum\limits_{r=0}^{\infty}B_{2r+2}^{(2n+2)}(l)B_{2r+4}^{(2n+2)}(l)\}  ] \frac{1}{2} \cos\alpha,
\end{eqnarray}
\begin{eqnarray}
&&A_{12} = \frac{Q}{2}[\{\frac{\pi}{2}\left(A_{1}^{(2n+1)}(l)\right)^{2}\nonumber\\&&+
\pi\sum\limits_{r=0}^{\infty}A_{2r+1}^{(2n+1)}(l)A_{2r+3}^{(2n+1)}(l)\}\nonumber\\&&+
 \{\pi\sum\limits_{r=0}^{\infty}B_{2r+2}^{(2n+2)}(l)B_{2r+4}^{(2n+2)}(l)\}  ] \frac{1}{2} \sin\alpha,
\end{eqnarray}
\begin{eqnarray}
&&A_{13} = \frac{Q}{2}[\{\frac{\pi}{2}\left(A_{1}^{(2n+1)}(l)\right)^{2}\nonumber\\&&+
 \pi\sum\limits_{r=0}^{\infty}A_{2r+1}^{(2n+1)}(l)A_{2r+3}^{(2n+1)}(l)\}\nonumber\\&&-
 \{\pi\sum\limits_{r=0}^{\infty}B_{2r+2}^{(2n+2)}(l)B_{2r+4}^{(2n+2)}(l)\}  ] \frac{1}{2} \cos\alpha,
\end{eqnarray}
\begin{eqnarray}
&&A_{14} = \frac{Q}{2}[\{\frac{\pi}{2}\left(A_{1}^{(2n+1)}(l)\right)^{2}\nonumber\\&&+
 \pi\sum\limits_{r=0}^{\infty}A_{2r+1}^{(2n+1)}(l)A_{2r+3}^{(2n+1)}(l)\}\nonumber\\&&-
 \{\pi\sum\limits_{r=0}^{\infty}B_{2r+2}^{(2n+2)}(l)B_{2r+4}^{(2n+2)}(l)\}  ] \frac{1}{2} \sin\alpha,
\end{eqnarray}
\begin{eqnarray}
&&A_{21} = \frac{Q}{2}[\{\frac{\pi}{2}\left(A_{1}^{(2n+1)}(l)\right)^{2}\nonumber\\&&+
 \pi\sum\limits_{r=0}^{\infty}A_{2r+1}^{(2n+1)}(l)A_{2r+3}^{(2n+1)}(l)\}\nonumber\\&&+
 \{\pi\sum\limits_{r=0}^{\infty}B_{2r+2}^{(2n+2)}(l)B_{2r+4}^{(2n+2)}(l)\}  ] \frac{1}{2} \sin\alpha,
\end{eqnarray}
\begin{eqnarray}
&&A_{22}=E_n(l)-\omega_0/2+\frac{Q}{2}[\{\frac{\pi}{2}\left(A_{1}^{(2n+1)}(l)\right)^{2}\nonumber\\&&+
 \pi\sum\limits_{r=0}^{\infty}A_{2r+1}^{(2n+1)}(l)A_{2r+3}^{(2n+1)}(l)\}+\{\frac{\pi}{2}\left(-B_{1}^{(2n+1)}(l)\right)^{2}\nonumber\\&&+
 \pi\sum\limits_{r=0}^{\infty}B_{2r+2}^{(2n+2)}(l)B_{2r+4}^{(2n+2)}(l)\}  ]\frac{1}{2} (-\cos\alpha) ,
\end{eqnarray}

\begin{eqnarray}
&&A_{23}=\frac{Q}{2}[\{\frac{\pi}{2}\left(A_{1}^{(2n+1)}(l)\right)^{2}\nonumber\\&&+
 \pi\sum\limits_{r=0}^{\infty}A_{2r+1}^{(2n+1)}(l)A_{2r+3}^{(2n+1)}(l)\}\nonumber\\&&-
 \{\pi\sum\limits_{r=0}^{\infty}B_{2r+2}^{(2n+2)}(l)B_{2r+4}^{(2n+2)}(l)\}  ]\frac{1}{2} \sin\alpha ,
\end{eqnarray}
\begin{eqnarray}
&&A_{24}=\frac{Q}{2}[\{\frac{\pi}{2}\left(A_{1}^{(2n+1)}(l)\right)^{2}\nonumber\\&&+
 \pi\sum\limits_{r=0}^{\infty}A_{2r+1}^{(2n+1)}(l)A_{2r+3}^{(2n+1)}(l)\}\nonumber\\&&-
 \{\pi\sum\limits_{r=0}^{\infty}B_{2r+2}^{(2n+2)}(l)B_{2r+4}^{(2n+2)}(l)\}  ]\frac{1}{2} (-\cos\alpha) ,\nonumber\\
\end{eqnarray}
\begin{eqnarray}
&&A_{31}=\frac{Q}{2}[\{\frac{\pi}{2}\left(A_{1}^{(2n+1)}(l)\right)^{2}\nonumber\\&&+
 \pi\sum\limits_{r=0}^{\infty}A_{2r+1}^{(2n+1)}(l)A_{2r+3}^{(2n+1)}(l)\}\nonumber\\&&-
\{ \pi\sum\limits_{r=0}^{\infty}B_{2r+2}^{(2n+2)}(l)B_{2r+4}^{(2n+2)}(l)\}  ]\frac{1}{2} \cos\alpha ,
\end{eqnarray}
\begin{eqnarray}
&&A_{32}=\frac{Q}{2}[\{\frac{\pi}{2}\left(A_{1}^{(2n+1)}(l)\right)^{2}\nonumber\\&&+
 \pi\sum\limits_{r=0}^{\infty}A_{2r+1}^{(2n+1)}(l)A_{2r+3}^{(2n+1)}(l)\}\nonumber\\&&-
 \{\pi\sum\limits_{r=0}^{\infty}B_{2r+2}^{(2n+2)}(l)B_{2r+4}^{(2n+2)}(l)\}  ]\frac{1}{2} \sin\alpha ,
\end{eqnarray}

\begin{eqnarray}
&&A_{33}=E_n(l)+\omega_0/2+\frac{Q}{2}[\{\frac{\pi}{2}\left(A_{1}^{(2n+1)}(l)\right)^{2}\nonumber\\&&+
 \pi\sum\limits_{r=0}^{\infty}A_{2r+1}^{(2n+1)}(l)A_{2r+3}^{(2n+1)}(l)\}\nonumber\\&&+
\{ \pi\sum\limits_{r=0}^{\infty}B_{2r+2}^{(2n+2)}(l)B_{2r+4}^{(2n+2)}(l)\}  ] \frac{1}{2} \cos\alpha ,
\end{eqnarray}
\begin{eqnarray}
&&A_{34}=\frac{Q}{2}[\{\frac{\pi}{2}\left(A_{1}^{(2n+1)}(l)\right)^{2}\nonumber\\&&+
 \pi\sum\limits_{r=0}^{\infty}A_{2r+1}^{(2n+1)}(l)A_{2r+3}^{(2n+1)}(l)\}\nonumber\\&&+
\{ \pi\sum\limits_{r=0}^{\infty}B_{2r+2}^{(2n+2)}(l)B_{2r+3}^{(2n+1)}(l)\}  ]\frac{1}{2} \sin\alpha ,
\end{eqnarray}
\begin{eqnarray}
&&A_{41}=\frac{Q}{2}[\{\frac{\pi}{2}\left(A_{1}^{(2n+1)}(l)\right)^{2}\nonumber\\&&+
 \pi\sum\limits_{r=0}^{\infty}A_{2r+1}^{(2n+1)}(l)A_{2r+3}^{(2n+1)}(l)\}\nonumber\\&&-
 \{\pi\sum\limits_{r=0}^{\infty}B_{2r+2}^{(2n+2)}(l)B_{2r+4}^{(2n+2)}(l)\}  ]\frac{1}{2} \sin\alpha ,
\end{eqnarray}
\begin{eqnarray}
&&A_{42}=\frac{Q}{2}[\{\frac{\pi}{2}\left(A_{1}^{(2n+1)}(l)\right)^{2}\nonumber\\&&+
 \pi\sum\limits_{r=0}^{\infty}A_{2r+1}^{(2n+1)}(l)A_{2r+3}^{(2n+1)}(l)\}\nonumber\\&&-
 \{\pi\sum\limits_{r=0}^{\infty}B_{2r+2}^{(2n+2)}(l)B_{2r+4}^{(2n+2)}(l)\}  ]\frac{1}{2} (-\cos\alpha) ,\nonumber
\end{eqnarray}
\begin{eqnarray}
&&A_{43}=\frac{Q}{2}[\{\frac{\pi}{2}\left(A_{1}^{(2n+1)}(l)\right)^{2}\nonumber\\&+&
 \pi\sum\limits_{r=0}^{\infty}A_{2r+1}^{(2n+1)}(l)A_{2r+3}^{(2n+1)}(l)\}\nonumber\\&&+
\{ \pi\sum\limits_{r=0}^{\infty}B_{2r+2}^{(2n+2)}(l)B_{2r+4}^{(2n+2)}(l)\}  ]\frac{1}{2} \sin\alpha ,
\end{eqnarray}
\begin{eqnarray}
&&A_{44}= E_n(l)-\omega_0/2+\frac{Q}{2}[\{\frac{\pi}{2}\left(A_{1}^{(2n+1)}(l)\right)^{2}\nonumber\\&&+
 \pi\sum\limits_{r=0}^{\infty}A_{2r+1}^{(2n+1)}(l)A_{2r+3}^{(2n+1)}(l)\}\nonumber\\&&+
\{ \pi\sum\limits_{r=0}^{\infty}B_{2r+2}^{(2n+2)}(l)B_{2r+4}^{(2n+2)}(l)\}  ] \frac{1}{2} (-\cos\alpha) ,\nonumber\\
\end{eqnarray}

\begin{widetext}
\section{Eigenvectors in $G_-$ region}
\label{appendix4}
Eigenvectors of Hamiltonian corresponding to $G_{-}$ region  in Eq. (\ref{ham_G-}) is given as:-

\begin{eqnarray}
&&\frac{(b_{1} - c_{1} \pm \sqrt{(b_{1} - c_{1})^2 + (d_{1} - e_{1})^2} )}{ \sqrt{
  2(d_{1}-e_{1})^2 \pm 2 (b_{1} - c_{1} \pm \sqrt{(b_{1} - c_{1})^2 + (d_{1} - e_{1})^2} )^2
   }}, -\frac{(d_{1}-e_{1})}{\sqrt{
  2(d_{1}-e_{1})^2 \pm 2 (b_{1} - c_{1} \pm \sqrt{(b_{1} - c_{1})^2 + (d_{1} - e{1})^2} )^2
   }},\nonumber \\
   && \frac{(-b_{1} + c_{1} \mp \sqrt{(b_{1} - c_{1})^2 + (d_{1} - e_{1})^2} )}{ \sqrt{
  2(d_{1}-e_{1})^2 \pm 2 (b_{1} - c_{1} \pm \sqrt{(b_{1} - c_{1})^2 + (d_{1} - e_{1})^2} )^2
   }},\frac{(d_{1}-e_{1})}{\sqrt{
  2(d_{1}-e_{1})^2 \pm 2 (b_{1} - c_{1} \pm \sqrt{(b_{1} - c_{1})^2 + (d_{1} - e_{1})^2} )^2
   }}
  \end{eqnarray}
\begin{eqnarray}
&&\frac{(b_{1} + c_{1} \pm \sqrt{(b_{1} + c_{1})^2 + (d_{1} + e_{1})^2} )}{ \sqrt{
  2(_{1}+e_{1})^2 + 2 (b_{1} + c_{1} \pm \sqrt{(b_{1} + c_{1})^2 + (d_{1} + e_{1})^2} )^2
   }}, \frac{(d_{1}+e_{1})}{\sqrt{
  2(d_{1}+e_{1})^2 + 2 (b_{1} + c_{1} \pm \sqrt{(b_{1} + c_{1})^2 + (d_{1} + e_{1})^2} )^2
   }}\nonumber \\,&& \frac{(b_{1} + c_{1} \pm \sqrt{(b_{1} + c_{1})^2 + (d_{1} + e_{1})^2} )}{ \sqrt{
  2(d_{1}+e_{1})^2 + 2 (b_{1} + c_{1} \pm \sqrt{(b_{1} + c_{1})^2 + (d_{1} + e_{1})^2} )^2
   }},\frac{(d_{1}+e_{1})}{\sqrt{
  2(d_{1}+e_{1})^2 + 2 (b_{1} + c_{1} \pm \sqrt{(b_{1} + c_{1})^2 + (d_{1} + e_{1})^2} )^2
   }}
 \end{eqnarray}
\end{widetext}
\section{Coefficients of density matrix in $G_-$ region}
\label{appendix5}
\begin{widetext}
\begin{eqnarray}
\zeta_{1}(t)&=&e^{-i a_{1} t}\bigg\{\frac{1}{2}\cos(t\lambda_1)+\frac{1}{2}\cos(t\lambda_2)-
 \frac{1}{2} i b_1 \bigg(\frac{\sin\big(t\lambda_1\big)}{\lambda_1}+\frac{\sin\big(t\lambda_2\big)}{\lambda_2}\bigg)
+\frac{1}{2} i c_{1}\bigg(\frac{\sin\big(t\lambda_1\big)}{\lambda_1}-\frac{\sin\big(t\lambda_2\big)}{\lambda_2}\bigg)\bigg\},
\end{eqnarray}
\begin{eqnarray}
\zeta_{2}(t)&=&e^{-i a_{1} t}\bigg\{-\frac{1}{2}ie_1\bigg(\frac{\sin\big(t\lambda_1\big)}{\lambda_1}+\frac{\sin\big(t\lambda_2\big)}{\lambda_2}\bigg)+\frac{1}{2}id_1\bigg(\frac{\sin\big(t\lambda_1\big)}{\lambda_1}-\frac{\sin\big(t\lambda_2\big)}{\lambda_2}\bigg)\bigg\},
\end{eqnarray}
\begin{eqnarray}
\zeta_{3}(t)=e^{-i a_{1} t}\bigg\{-\frac{1}{2}i c_1\bigg(\frac{\sin\big(t\lambda_1\big)}{\lambda_1}+\frac{\sin\big(t\lambda_2\big)}{\lambda_2}\bigg)+
\frac{1}{2} \cos\big(t\lambda_2\big)-\frac{1}{2} \cos\big(t\lambda_1\big)
+\frac{1}{2}i b_{1}\bigg(\frac{\sin\big(t\lambda_1\big)}{\lambda_1}-\frac{\sin\big(t\lambda_2\big)}{\lambda_2}\bigg)\bigg\},
\end{eqnarray}
\begin{eqnarray}
\zeta_{4}(t)&=&e^{-i a_{1} t}\bigg\{-\frac{1}{2}id_1\bigg(\frac{\sin\big(t\lambda_1\big)}{\lambda_1}+\frac{\sin\big(t\lambda_2\big)}{\lambda_2}\bigg)+
\frac{1}{2}i e_{1}\bigg(\frac{\sin\big(t\lambda_1\big)}{\lambda_1}-\frac{\sin\big(t\lambda_2\big)}{\lambda_2}\bigg)\bigg\}.
\end{eqnarray}
\end{widetext}
The following notations are used:
\begin{eqnarray}
\lambda_{1,2}=\sqrt{(b_1 \mp c_1)^2+(d_1 \mp e_1^2)^2 }.
\end{eqnarray}
\section{Calculation of $d|c_1(t)|/dt$}
\label{appendix6}
\begin{widetext}
\begin{eqnarray}
&&F=\frac{d{\vert c_{1}(t)\vert}}{dt}=\frac{-\tau}{2}\vert c_{1}(t)\vert+\frac{\vert c_{1}(0)\vert^{2}e^{-\tau t}}{\vert c_{1}(t)\vert}\Bigg(\Big(\cosh{\Big(\frac{\kappa' t}{2}\Big)}\cos{\Big(\frac{\kappa'' t}{2}\Big)}+\frac{\kappa_1}{\vert\kappa_{x}\vert^2}\sinh{\Big(\frac{\kappa' t}{2}\Big)}\cos{\Big(\frac{\kappa'' t}{2}\Big)}-\frac{\kappa_2}{\vert\kappa_{x}\vert^2}\cosh{\Big(\frac{\kappa' t}{2}\Big)}\sin{\Big(\frac{\kappa'' t}{2}\Big)}\Big)\nonumber\\&&\Big(\frac{\kappa'}{2}\Big(\sinh{\Big(\frac{\kappa' t}{2}\Big)}\cos{\Big(\frac{\kappa'' t}{2}\Big)}+\frac{\kappa_1}{\vert\kappa_{x}\vert^2}\cosh{\Big(\frac{\kappa' t}{2}\Big)}\cos{\Big(\frac{\kappa'' t}{2}\Big)}-\frac{\kappa_2}{\vert\kappa_{x}\vert^2}\sinh{\Big(\frac{\kappa' t}{2}\Big)}\sin{\Big(\frac{\kappa'' t}{2}\Big)}\Big)+\frac{\kappa''}{2}\Big(-\cosh{\Big(\frac{\kappa' t}{2}\Big)}\sin{\Big(\frac{\kappa'' t}{2}\Big)}\nonumber\\&&-\frac{\kappa_1}{\vert\kappa_{x}\vert^2}\sinh{\Big(\frac{\kappa' t}{2}\Big)}\sin{\Big(\frac{\kappa'' t}{2}\Big)}-\frac{\kappa_2}{\vert\kappa_{x}\vert^2}\cosh{\Big(\frac{\kappa' t}{2}\Big)}\cos{\Big(\frac{\kappa'' t}{2}\Big)}\Big)\Big)+  \Big(\sinh{\Big(\frac{\kappa' t}{2}\Big)}\sin{\Big(\frac{\kappa'' t}{2}\Big)}+\frac{\kappa_1}{\vert\kappa_{x}\vert^2}\cosh{\Big(\frac{\kappa' t}{2}\Big)}\sin{\Big(\frac{\kappa'' t}{2}\Big)}\nonumber\\&&+\frac{\kappa_2}{\vert\kappa_{x}\vert^2}\sinh{\Big(\frac{\kappa' t}{2}\Big)}\cos{\Big(\frac{\kappa'' t}{2}\Big)}\Big)\Big(\frac{\kappa'}{2}\Big(\cosh{\Big(\frac{\kappa' t}{2}\Big)}\sin{\Big(\frac{\kappa'' t}{2}\Big)}+\frac{\kappa_1}{\vert\kappa_{x}\vert^2}\sinh{\Big(\frac{\kappa' t}{2}\Big)}\sin{\Big(\frac{\kappa'' t}{2}\Big)}+\frac{\kappa_2}{\vert\kappa_{x}\vert^2}\cosh{\Big(\frac{\kappa' t}{2}\Big)}\cos{\Big(\frac{\kappa'' t}{2}\Big)}\Big)\nonumber\\&&+\frac{\kappa''}{2}\Big(-\sinh{\Big(\frac{\kappa' t}{2}\Big)}\cos{\Big(\frac{\kappa'' t}{2}\Big)}-\frac{\kappa_1}{\vert\kappa_{x}\vert^2}\cosh{\Big(\frac{\kappa' t}{2}\Big)}\cos{\Big(\frac{\kappa'' t}{2}\Big)}+\frac{\kappa_2}{\vert\kappa_{x}\vert^2}\sinh{\Big(\frac{\kappa' t}{2}\Big)}\sin{\Big(\frac{\kappa'' t}{2}\Big)}\Big)\Big)\Bigg)\nonumber\\
\end{eqnarray}
Where $\kappa'=\sqrt{\frac{\sqrt{(\tau^2-2Ng-\delta)^2+(2\tau\delta)^2}+(\tau^2-2Ng-\delta)}{2}}$ \Bigg($\kappa''=\sqrt{\frac{\sqrt{(\tau^2-2Ng-\delta)^2+(2\tau\delta)^2}-(\tau^2-2Ng-\delta)}{2}}$\Bigg) are real (imaginary) part of $\kappa_{x}$. $\kappa_{1}= (\tau \kappa'+\delta \kappa'' )$, $\kappa_{2}= ( \delta\kappa'- \tau\kappa'' )$.
\end{widetext}
\section{ Reduced density matrix $\rho_s(t)$}
\label{appendix7}
\begin{widetext}
The reduced density matrix of the system in $G_{0}$ region coupled to N bosonic reservoirs can be obtained from the total state $\vert \Psi_{0}(0)\rangle=\vert \psi_{0}(0)\rangle\otimes\prod_{n=1}^{N}\vert \bar{0}\rangle$ which evolves in time as:
\begin{eqnarray}
\vert\Psi(t)\rangle=[c_{0}(0)\vert0\rangle+c_{1}(t)\vert1\rangle]\otimes\prod_{n=1}^{N}\vert\Bar{0}\rangle_{n,r}+\vert0\rangle\otimes\sum_{n=1}^{N}\sum_{k}c_{n,k}(t)\vert1_{k}\rangle_{n,r}.
\end{eqnarray}
\begin{eqnarray}
\vert\Bar{0}\rangle_{n,r}=\prod_{k=1}\vert{0_{k}}\rangle_{n,r},\prod_{k=1}\langle{0_{k}} \vert 0_{k}\rangle_{n,r}=1,            \prod_{n=1}^{N}\langle \bar{0} \vert \bar{0}\rangle_{n,r}=1,
\end{eqnarray}
Normalization condition is:
\begin{eqnarray}
\langle \Psi(t)\vert \Psi(t)\rangle&=&([c_{0}^{*}(0)\langle 0\vert+c_{1}^{*}(t)\langle1\vert]\otimes\prod_{n'=1}^{N}\langle\Bar{0}\vert_{n,r}+\langle0\vert \otimes \sum_{n'=1}^{N}\sum_{k'}c_{n,k}^{*}(t)\langle1_{k'}\vert_{n',r})[c_{0}(0)\vert0\rangle+c_{1}(t)\vert1\rangle]\otimes\prod_{n=1}^{N}\vert\Bar{0}\rangle_{n,r}\nonumber\\&+&\vert0\rangle \otimes \sum_{n=1}^{N}\sum_{k}c_{n,k}(t)\vert1_{k}\rangle_{n,r}
\end{eqnarray}
which comes out to be
\begin{eqnarray}
\vert c_{0}(0)\vert^2+\vert c_{1}(t)\vert^2+ \sum_{n=1}^{N}\sum_{k}\vert c_{n,k}(t)\vert^2=1.
\end{eqnarray}
The state of the system at any time $t$ can be written in the form
\begin{eqnarray}
\vert\Psi(t)\rangle=\Big[c_{0}(0)\otimes \prod_{n=1}^{N}\prod_{k=1}\vert {0_{k}}\rangle_{n,r}+\mathcal{I}\otimes \sum_{n=1}^{N}\sum_{k}c_{n,k}(t)\vert1_{k}\rangle_{n,r}\Big]\otimes\vert{0}\rangle +c_{1}(t)\vert1\rangle \otimes \prod_{n=1}^{N}\prod_{k=1}\vert {0_{k}}\rangle_{n,r}
\end{eqnarray}
and corresponding density matrix can be written as
\begin{eqnarray}
&&\rho(t)=\vert\Psi(t)\rangle\langle\Psi(t)\vert\nonumber\\&&=\Big[c_{0}(0)\otimes \prod_{n=1}^{N}\prod_{k=1}\vert {0_{k}}\rangle_{n,r}+\mathcal{I}\otimes \sum_{n=1}^{N}\sum_{k}c_{n,k}(t)\vert1_{k}\rangle_{n,r}\Big] \Big[c_{0}^{*}(0)\otimes \prod_{n'=1}^{N}\prod_{k'=1}\langle {0_{k'}}\vert_{n',r}+\mathcal{I}\otimes \sum_{n'=1}^{N}\sum_{k'}c_{n',k'}^{*}(t)\langle 1_{k'}\vert_{n',r}\Big]\vert0\rangle\langle 0 \vert \nonumber\\&&+\Big[c_{0}(0)\otimes \prod_{n=1}^{N}\prod_{k=1}\vert {0_{k}}\rangle_{n,r}+\mathcal{I}\otimes \sum_{n=1}^{N}\sum_{k}c_{n,k}(t)\vert1_{k}\rangle_{n,r}\Big]c_{1}^{*}(t)\otimes \prod_{n'}^{N}\prod_{k'=1}\langle0_{k'}\vert_{n',r} \vert0\rangle \langle 1 \vert+c_{1}(t)\otimes\prod_{n=1}^{N}\prod_{k=1}\vert0_{k}\rangle_{n,r}\nonumber\\&& \Big[c_{0}^{*}(0)\otimes \prod_{n=1}^{N}\prod_{k'=1}\langle {0_{k'}}\vert_{n',r}+\mathcal{I}\otimes \sum_{n'=1}^{N}\sum_{k'}c_{n',k'}^{*}(t)\langle 1_{k'}\vert_{n',r}\Big]\vert1\rangle\langle0\vert+\vert c_{1}(t)\vert^2\otimes\prod_{n=1}^{N}\prod_{k=1}\vert0_{k}\rangle_{n,r}\prod_{n'=1}^{N}\prod_{k'=1}\langle0_{k'}\vert_{n',r}\vert 1\rangle \langle 1 \vert,
\end{eqnarray}
Now, tracing out the reservoir part, we get reduced density matrix in terms of Probability amplitude $c_{0}(0),c_{1}(t)$ as:
\begin{eqnarray}
\rho_{s}(t)=tr_{r}(\vert\Psi(t)\rangle\langle\Psi(t)\vert)=(1-\vert  c_{1}(t)\vert^2)\vert0\rangle\langle0\vert+c_{0}(0)c_{1}^{*}(t)\vert0\rangle\langle1\vert+c_{1}(t)c_{0}^{*}(0)\vert1\rangle\langle0\vert+\vert c_{1}(t)\vert^2\vert1\rangle\langle1\vert.
\end{eqnarray}
\end{widetext}
\section{Eigenvalues and eigenvectors related to section \ref{unitary_coherence}}
\label{appendix_F}
\begin{widetext}
Eigenvalues corresponding to Hamiltonian Eq.\ref{Hamiltonian_after_quench}
\begin{eqnarray} 
&& E_{\hat{H},1,n}(l)=\frac{1}{2}\bigg(E_{1,n}(l)+E_{2,n}(l)+\sqrt{4Y^{2}+
(E_{1,n}(l)-E_{2,n}(l)+2J)^{2}}\bigg),\nonumber\\
&& E_{\hat{H},2,n}(l)=\frac{1}{2}\bigg(E_{1,n}(l)+E_{2,n}(l)-\sqrt{4Y^{2}+
(E_{1,n}(l)-E_{2,n}(l)+2J)^{2}}\bigg).
\end{eqnarray}

The coefficient of eigenvectors represented in Eq.\ref{are the eigenfunctions of2} are:
\begin{eqnarray}
\zeta_{1}=\frac{L}{\sqrt{L^{2}+1}}=\sin\Theta, {\rm and}\ \zeta_{2}=\frac{1}{\sqrt{L^{2}+1}}=\cos\Theta,
\end{eqnarray}
where,
\begin{eqnarray}
L=\frac{E_{1,n}(l)-E_{2,n}(l)+2J}{2Y}+\frac{\sqrt{4Y^{2}+
(E_{1,n}(l)-E_{2,n}(l)+2J)^{2}}}{2Y},
\end{eqnarray}
and
\begin{eqnarray}
J=-\Delta \omega_{0}(\alpha_{1}^{2}-\beta_{1}^{2}), \ Y=-2\Delta\omega_{0}\alpha_{1}\beta_{1}
\end{eqnarray}
\begin{eqnarray}
&&E_{1,n}(l)=a(l)+\sqrt{\frac{1}{4}\omega_{0}^{2}+\frac{1}{4}Q^{2}\langle ce_{n}(l)|\cos2\varphi|ce_{n}(l)\rangle^{2}+\frac{1}{2}\omega_{0}Q\cos\alpha\langle ce_{n}(l)|\cos2\varphi|ce_{n}(l)\rangle},\\
&&E_{2,n}(l)=a(l)-\sqrt{\frac{1}{4}\omega_{0}^{2}+\frac{1}{4}Q^{2}\langle ce_{n}(l)|\cos2\varphi|ce_{n}(l)\rangle^{2}+\frac{1}{2}\omega_{0}Q\cos\alpha\langle ce_{n}(l)|\cos2\varphi|ce_{n}(l)\rangle},
\end{eqnarray}
The coefficient of eigenvectors of $\rho(t)$ represented in Eq.\ref{are the eigenfunctions of3} are:
\begin{eqnarray}
\nu_{1}=-\frac{\tan(\Theta)\exp(-i\omega_{12}t)}{\sqrt{\tan(\Theta)^{2}+1}},\\
\nu_{2}=\frac{1}{\sqrt{\tan(\Theta)^{2}+1}},
\end{eqnarray}
and the eigenvalues of $\rho(t)$ are
\begin{eqnarray}
E_{1}(\rho(t))=p_{1},\ E_{2}(\rho(t))=p_{2}.
\end{eqnarray}

\begin{eqnarray}
&&\lambda=\frac{2}{Q\sin\alpha\langle ce_{n}(l)|\cos2\varphi|ce_{n}(l)\rangle}\bigg\{\frac{1}{2}\omega_{0}+\frac{1}{2}Q\cos\alpha\langle ce_{n}(l)|\cos2\varphi|ce_{n}(l)\rangle+\nonumber\\ &&\sqrt{\frac{1}{4}\omega_{0}^{2}+\frac{1}{4}Q^{2}\langle ce_{n}(l)|\cos2\varphi|ce_{n}(l)\rangle^{2}+\frac{1}{2}\omega_{0}Q\cos\alpha\langle ce_{n}(l)|\cos2\varphi|ce_{n}(l)\rangle}\bigg\}.
\end{eqnarray}

The value of the integral:

\begin{eqnarray}\label{through the Furrier representation}
&&\langle ce_{2n+1}(l)|V|ce_{2n+1}(l)\rangle=\Delta l\left\{\frac{\pi}{2}\left(A_{1}^{(2n+1)}(l)\right)^{2}+\pi\sum\limits_{r=0}^{\infty}A_{2r+1}^{(2n+1)}(l)A_{2r+3}^{(2n+1)}(l)\right\},\\
&&\langle ce_{2n}(l)|V|ce_{2n}(l)\rangle=\Delta l\left\{\pi A_{0}^{(2n)}(l)A_{2}^{(2n)}(l)+\pi\sum\limits_{r=0}^{\infty}A_{2r}^{(2n)}(l)A_{2r+2}^{(2n)}(l)\right\}.
\end{eqnarray}

Parameters for small barrier limit:

\begin{eqnarray}
&&E_{1,n}(l)=n^{2}+\sqrt{\frac{1}{4}\omega_{0}^{2}+\frac{1}{4}Q^{2}+\frac{1}{2}Q\omega_{0}\cos\alpha},\\
&&E_{2,n}(l)=n^{2}-\sqrt{\frac{1}{4}\omega_{0}^{2}+\frac{1}{4}Q^{2}+\frac{1}{2}Q\omega_{0}\cos\alpha},\\
&&\lambda=\frac{2}{Q\sin\alpha}\bigg\{\frac{1}{2}\omega_{0}+\frac{1}{2}Q\cos\alpha+ \sqrt{\frac{1}{4}\omega_{0}^{2}+\frac{1}{4}Q^{2}+\frac{1}{2}Q\omega_{0}\cos\alpha}\bigg\}.
\end{eqnarray}
\end{widetext}

\bibliographystyle{apsrev4-1}
\bibliography{nems}

\end{document}